\documentclass[lettersize,journal]{IEEEtran}
\usepackage{amsmath,amsfonts}

\usepackage{algorithm}
\usepackage{array}
\usepackage{textcomp}
\usepackage{stfloats}
\usepackage{url}
\usepackage{verbatim}
\usepackage{graphicx}
\usepackage{cite}
\usepackage{rotating}

\usepackage{adjustbox}
\usepackage{multirow}
\usepackage{color}
\usepackage{colortbl}
\usepackage{algpseudocode}
\usepackage{rotating}
\usepackage{tikz}
\usepackage{caption}
\usepackage{subcaption}

\usepackage{hyperref}
\hypersetup{
    colorlinks=false,
    linkcolor=blue,
    filecolor=green,      
    urlcolor=purple,
    }

\newcommand{\updated}[1]{\textcolor{black}{#1}}

\begin{document}

\title{\updated{Cognitive Load-based Affective Workload Allocation for Multi-human Multi-robot Teams}}

\author{Wonse~Jo,~
        Ruiqi~Wang,~\IEEEmembership{Graduate Student Member,~IEEE,}
        Baijian~Yang,~\IEEEmembership{Member,~IEEE,}
        Dan~Foti,~
        Mo~Rastgaar,~\IEEEmembership{Senior Member,~IEEE,}
        and~Byung-Cheol~Min,~\IEEEmembership{Senior Member,~IEEE}        
\IEEEcompsocitemizethanks{\IEEEcompsocthanksitem W.S. Jo, R.Q. Wang, B. Yang, B.-C. Min was with the Department of Computer and Information Technology, Purdue University, West Lafayette,
IN, 47906. E-mail: \{jow, wang5357, byang, minb\}@purdue.edu
\IEEEcompsocthanksitem D. Foti was with the Department of Psychological Sciences, Purdue University, West Lafayette,
IN, 47906. E-mail: foti@purdue.edu
\IEEEcompsocthanksitem M. Rastgaar was with the School of Engineering Technology, Purdue University, West Lafayette,
IN, 47906. E-mail: rastgaar@purdue.edu

}
\thanks{Manuscript received April 19, 2021; revised August 16, 2021.}}

\markboth{Journal of \LaTeX\ Class Files,~Vol.~14, No.~8, August~2021}%
{Shell \MakeLowercase{\textit{et al.}}: A Sample Article Using IEEEtran.cls for IEEE Journals}


\maketitle

\begin{abstract}
The interaction and collaboration between humans and multiple robots represent a novel field of research known as human multi-robot systems. 
Adequately designed systems within this field allow teams composed of both humans and robots to work together effectively on tasks such as monitoring, exploration, and search and rescue operations. 
This paper presents a deep reinforcement learning-based affective workload allocation controller specifically for multi-human multi-robot teams. 
The proposed controller can dynamically reallocate workloads based on the performance of the operators during collaborative missions with multi-robot systems. 
The operators' performances are evaluated through the scores of a self-reported questionnaire (i.e., subjective measurement) and the results of a deep learning-based cognitive workload prediction algorithm that uses physiological and behavioral data (i.e., objective measurement). 
To evaluate the effectiveness of the proposed controller, we conduct an exploratory user experiment with various allocation strategies. The user experiment uses a multi-human multi-robot CCTV monitoring task as an example and carry out comprehensive real-world experiments with 32 human subjects for both quantitative measurement and qualitative analysis. 
Our results demonstrate the performance and effectiveness of the proposed controller and highlight the importance of incorporating both subjective and objective measurements of the operators' cognitive workload as well as seeking consent for workload transitions, to enhance the performance of multi-human multi-robot teams.
\end{abstract}


\section{Introduction}
\label{sec:introduction}
\IEEEPARstart{A}{s} artificial intelligence continues to advance, multi-robot systems (MRS) are demonstrating consistent performance and precision that surpasses human ability in various large-scale operations, such as surveillance \cite{kolling2008multi} and, search and rescue \cite{luo2011multi}, and assembly \cite{sun2002adaptive}. However, compared to human's capabilities, MRS still has deficiencies in situational awareness (SA) when it comes to effectively handling complex task dynamics in the real world \cite{hoffman2004collaboration}. For example, adjusting the operation of an MRS in a timely manner in response to new missions and environmental changes can be challenging when operating the system for an extended period of time. This issue is currently mitigated by having a human operator participate in task execution, which improves efficiency. However, systems with many robots can produce excessively high cognitive workload (CWL) for a single human operator, making it difficult for them to track each robot's work. This can be addressed by having multiple operators in the loop to provide some level of supervision, resulting in a multi-human multi-robot (MH-MR) team. 

While incorporating human operators as the core of the decision-making process can significantly improve the system's SA and flexibility, it can also introduce more uncertainty and complexity. Human affective conditions such as CWL and emotion, as well as performance, are inconsistent and susceptible to internal or external factors \cite{kolb2022leveraging,hooey2017underpinnings,lyons2012human}. Thus, optimizing the performance of the entire MH-MR team, including optimal workload and workload allocation among multiple humans, is a crucial challenge. The system must monitor human affective states and reallocate workload accordingly, such as the number of robots to be supervised, to maintain each human in optimal interaction conditions with robots \cite{dahiya2022survey,barnes2015designing,feigh2014requirements}.

\begin{figure*}[t]
    \centering
    \includegraphics[width=1\linewidth]{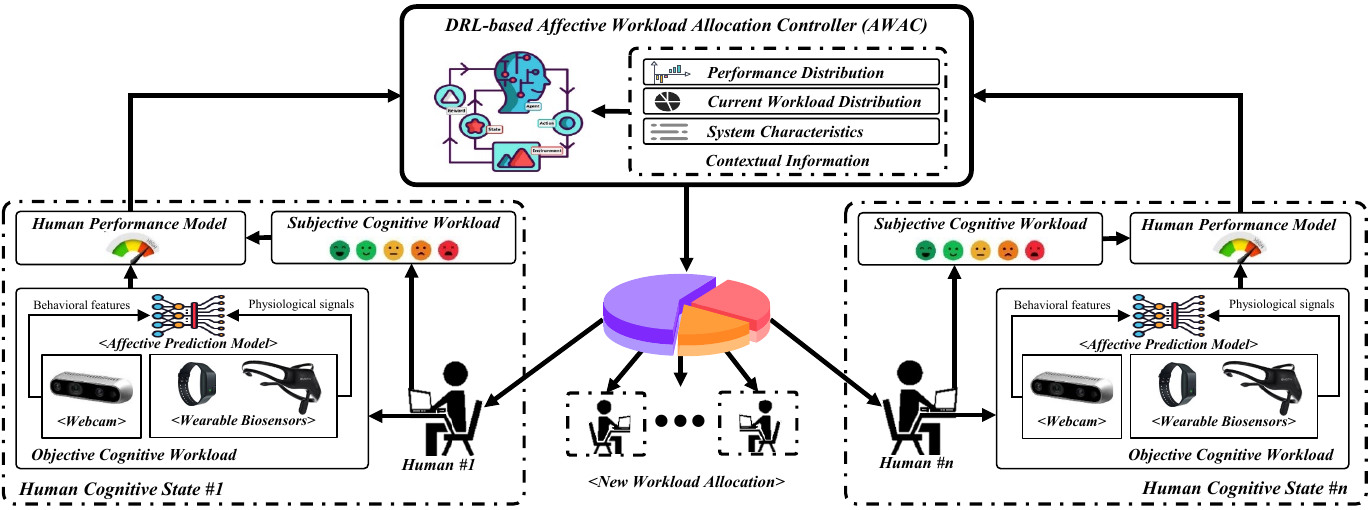}
    \caption{Conceptual illustration of the deep reinforcement learning (DRL)-based affective workload allocation controller (AWAC) for multi-human multi-robot (MH-MR) teams. More details can be found at Session \ref{sec:proposed_system} or supplementary website: \url{https://sites.google.com/view/affective-workload-allocation}.}
    \label{fig:AWA_concept}
\end{figure*}

While the workload allocation in MRS \cite{khamis2015multi,hardin2009using,miller2002playbook} and the scheduling of tasks in human-autonomy collaboration \cite{creech2021resource,gutzwiller2015human,tokadli2021evaluation} have been well studied, there has been limited research on the objective of optimally coordinating multiple humans in MH-MR team. In the limited literature available, existing works on workload allocation in MH-MR teams that take human factors into account are still in the preliminary stage and have several drawbacks. Most methods only consider human performance metrics or task difficulty as indicators reflecting human decision-making ability without considering individual human CWL \cite{ijtsma2019computational,talebpour2019adaptive}.

However, CWL, which measures the mental capacity required to complete tasks \cite{debie2019multimodal}, serves as a vital factor that influences the human ability to process environmental information and make decisions \cite{harriott2015mental,biondi2021overloaded,roy2020can,heard2018survey}. Therefore, neglecting this fundamental benchmark makes it difficult to imitate and encode unstructured human decision-making processes. For example, a decrease in human performance could result from both cognitive overload and underload on the task, making it challenging to determine the optimal task reallocation strategy (e.g., increasing or decreasing the current workload) without considering CWL.

Moreover, most existing approaches tend to build a model that encodes the relationship between system attributes, including contextual information and human factors, and the system performance, to serve as one-step rules for determining the optimal workload distribution \cite{fusaro2021integrated,mina2020adaptive}. However, building a valid and generalizable model is challenging due to the complexity of an MH-MR team and the individual differences between humans and task scenarios. Additionally, current workload allocation models are difficult to deploy in realistic MH-MR task scenarios as they mostly lack monitoring and assessment of contextual information and human states, and have barely been validated by real-world experiments.

To address these limitations (details in Section \ref{sec:background}), we propose a deep reinforcement learning (DRL)-based affective workload allocation controller (AWAC) for MH-MR teams, as illustrated in Fig. \ref{fig:AWA_concept}, 
\updated{in where the affective workload refers to changing the amount of the workload based on each human operator's task performance estimated from subjective and objective CWL measurements \cite{kalyuga2011cognitive,jo2024smart}. Thus, } 
the proposed controller can adaptively reallocate workloads based on the operators' performance during collaborative missions with the MRS. Operator performance is estimated by self-reporting questionnaire scores (i.e., subjective measurement) and the results of a deep learning (DL)-based CWL prediction algorithm using physiological and behavioral data (i.e., objective measurement). To evaluate our proposed system, we use a closed-circuit television (CCTV) monitoring task by a MH-MR team as an example and conduct extensive real-world team-based user experiments for quantitative measurement and qualitative analysis.

The main contributions of this paper are as follows:

\begin{itemize}
    \item We design a data-driven human performance model to estimate the human operator's mission performance from CWL measurements. It can be adapted to various applications by tuning parameters based on empirical experiments.
    \item We propose a DRL-based AWAC capable of adapting the distribution of workload in response to variations in human cognitive load and team performance.
    \item We design and conduct an extensive real-world user study in CCTV surveillance scenarios to validate the productivity and effectiveness of the proposed AWAC.
    \item We investigate and furnish insightful analysis of various workload allocation strategies for MH-MR teams.    
\end{itemize}

The paper is organized as follows. Section \ref{sec:background} presents the background and related works. Section \ref{sec:proposed_system} provides details of the proposed affective workload allocation controller. Section \ref{sec:design_user_exp} describes the design of a team-based monitoring task that was conducted to validate the proposed controller. In Section \ref{sec:result_analysis}, we present the results of the extensive user experiments and analyze the team performance of the proposed affective workload allocation system. Lastly, Section \ref{sec:discussion} discusses the findings of this study in depth, and Section \ref{sec:conclusion} concludes the proposed the affective workload allocation system.

\section{Related works} 
\label{sec:background}
In this section, we introduce the background of an overview of existing research on workload allocation in MH-MR Teams.

Workload allocation in MH-MR teams is a critical issue for practical human-robot interaction (HRI) applications. It is essential for mission completion and success, as it facilitates a clear division of workloads, increased efficiency and productivity, \updated{reduced workload} on individual operators, and improved adaptability to changing circumstances in the work environment \cite{harriott2015mental}. Assigning tasks to the most-suitable human operator helps avoid conflicts and enhances collaboration and communication between human and robotic agents \cite{mina2020adaptive}. With a well-defined workload allocation strategy, the MH-MR team can work cohesively toward achieving the assigned missions.

Previous research has examined human-in-the-loop systems, such as team organization, task scheduling \cite{chien2012scheduling}, and studies on SA in HRI \cite{drury2003awareness}. Workload allocation strategies based on human CWL \cite{kaber2006situation} have also been researched to control unmanned aerial vehicles \cite{prinzel2000closed,wilson2003operator}, and task performance and difficulty of human operators have been studied to reallocate workloads during missions \cite{parasuraman2007adaptive}.
However, these allocation strategies have typically prioritized system output while overlooking the needs of human operators. As a result, the majority of workload allocation research has been limited to applications in MH-MR teams in real-world settings.

According to \cite{roy2020can}, the utilization of human CWL is crucial in HRI applications for maximizing the entire performance, including productivity and effectiveness. This is because robotic or autonomous systems can adjust their workloads and control parameters (e.g., speed, control method) based on the human operator's condition, which can be affected by factors, \updated{such as personal reasons, negative emotions, and unusual environments.} Two primary methods are used to measure human CWL: subjective measurement and objective measurement. 

Subjective measurement, such as self-reported ratings or questionnaires, provides insight into the user's perceived level of CWLs related to factors such as motivation and prior experience. However, it is difficult to measure the operator's CWL in real-time, and there is a high possibility of bias or fake CWL, which is intentionally generated by human operators to reduce their workload. 

On the other hand, objective measurement, such as physiological measures, provides a more quantifiable assessment of CWL. They can reveal specific aspects of the interaction that are causing increased mental effort, such as complex visual displays or cognitively demanding tasks. Some studies in the affective computing and HRI fields have used physiological sensors to reflect the operator's CWL or emotional states \cite{9223531}. Other studies have utilized behavioral data, such as head pose, eye blinking and gazing, as well as dynamic movement of the input interfaces, to estimate affective states \cite{narayanan2020proxemo,sherry2002dynamic}. 
However, systems using objective measurement are vulnerable to malfunctioning physiological sensors or behavioral monitoring devices, which can result in inaccurate measurements of human CWL and impair the overall system's performance.

By incorporating both subjective and objective measurements, the limitations of each method can be mitigated, resulting in a more comprehensive understanding of the human's CWL in HRI applications. This, in turn, enables better-informed design decisions and an enhanced overall user experience. 


Subjective and objective CWL measurements, when used in conjunction with reinforcement learning (RL), can serve as a workload allocation controller for MH-MR teams. RL optimizes workloads based on both measurements, allowing for real-time allocation of tasks among multiple humans and robots to achieve desired outcomes. This results in a more effective and efficient system that balances the needs of both humans and robots. Furthermore, RL allows for adaptation to changing conditions, which is crucial in dynamic and uncertain human-robot interaction scenarios. Despite its potential benefits, there is limited understanding, and fewer studies that have explored the use of both measurements and RL approaches for MH-MR applications in real-world settings. 

Previous research has predominantly focused on the use of RL for workload allocation in human-robot interaction scenarios \cite{zhang2022reinforcement, ghadirzadeh2016sensorimotor}. These studies aimed to balance the workload between humans and robots in collaborative tasks while considering factors such as user satisfaction, task efficiency, and cognitive load. 
Zhang \textit{et. al} \cite{zhang2022reinforcement}, for instance, proposed an algorithm to optimize task allocation in complex assembly operations. The algorithm was evaluated through a virtual assembly of an alternator and showed great potential in reducing human workload and improving efficiency in human-robot collaboration tasks.
Lim \textit{et. al} \cite{lim2021adaptive} presented a human-machine interface and interaction system to support adaptive automation in unmanned aircraft systems. This system uses a network of physiological sensors and machine learning models to infer the user's CWL in single human operator and MRS scenarios, where the human operator is responsible for coordinating the tasks of multiple UAVs.
Ghadirzadeh \textit{et. al} \cite{ghadirzadeh2016sensorimotor} developed a data-efficient RL framework for modeling physical human-robot collaborations that enables the robot to learn how to collaborate with a human operator. The framework reduces uncertainty in the interaction using Gaussian processes, and optimal action selection is ensured through Bayesian optimization. 

However, these approaches have primarily focused on optimizing task allocation in single human-robot interaction scenarios, which limits the applicability of workload allocation research to real-world collaborations involving multiple robots and human operators. Additionally, these approaches have not comprehensively considered the various conditions of the human operator and have not taken into account the human operator's CWL in collaborative missions with multiple robots. Furthermore, the narrow focus on specific learning models limits their ability to adapt to unexpected situations.

\section{Affective Workload Allocation Controller}
\label{sec:proposed_system}
In this section, we introduce our AWAC for MH-MR teams. The AWAC aims to enable human operators to collaborate more effectively with multi-robot systems and teammates by intelligently assigning appropriate workloads based on both subjective and objective measurements of the CWL of the operator and their teammates. 
For example, if one operator has a high CWL, the proposed AWAC will assign more work to other operators to balance the workload. The AWAC is designed to improve the efficiency and effectiveness of MH-MR teams by mitigating the impact of CWL on task performance. 
Additionally, there is a supplementary website including more details of the AWAC at \url{https://sites.google.com/view/affective-workload-allocation}.

\begin{figure}[t]
    \centering
    \includegraphics[width=0.9\linewidth]{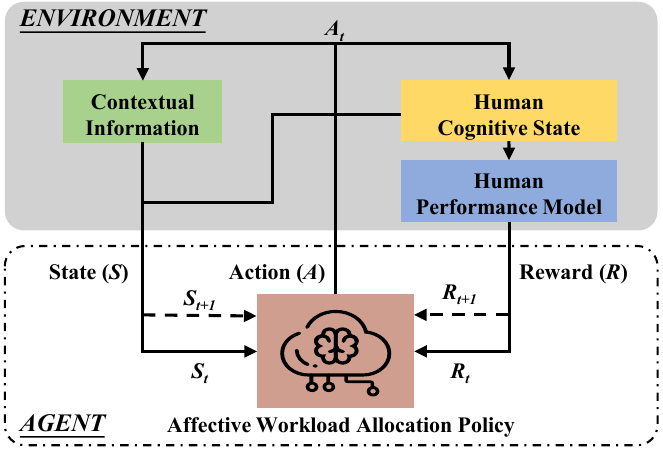}
    \caption{A learning diagram of the proposed DRL model.}
    \label{fig:drl_diagram_model}
\end{figure}

\subsection{Problem Formulation}
Although having knowledge of contextual information and human conditions (e.g., CWL) can serve as a decision-making foundation for the workload allocation problem, it is still a challenging task to determine the ideal workload distribution, given the complex nature of MH-MR teams. Unlike existing model-based approaches that rely on a single-step allocation rule model, we address the workload allocation in MH-MR teams as a partially observable Markov decision process (POMDP) and apply DRL to find the optimal solution. The partial observability arises from the limited ability of any Affective Prediction Model (APM) to accurately and completely predict continuous human conditions during a task. The proposed POMDP for the affective workload allocation in MH-MR teams is defined as a tuple $(\mathcal{H}, \mathcal{W}, \mathcal{S}, \mathcal{O},\mathcal{A}, \mathcal{T}, \mathcal{R}, \gamma)$, where:

\begin{itemize}
    \item $\mathcal{H}$ := \{${h_1}$, \dots, ${h_n}$\} is a finite set of $n$ human agents.
    \item $\mathcal{W}$ := \{${w_1}$, \dots, ${w_i}$\} is a finite set of $i$ tasks, e.g., the number of robots to be supervised, to be assigned to $n$ human agents.
    \item $\mathcal{S}$ := \{${S^s}$ $\times$ ${S^o}$\} is the joint human state observed, including subjective CWL obtained by the self-reporting questionnaire, ${S^s}$ := \{${s^s_1}$ $\times$ \dots $\times$ ${s^s_n}$\}, and objective CWL assessed by the APM, ${S^o}$ := \{${s^o_1}$ $\times$ \dots $\times$ ${s^o_n}$\}, of the $n$ human agents.
    \item $\mathcal{O}$ := \{${o_1}$ $\times$ \dots $\times$ ${o_q}$\} is the joint observation of the contextual information, e.g., current workload distribution, performance metrics, and other important system characteristics.
    \item $\mathcal{A}$ := \{${a_1}$ $\times$ \dots $\times$ ${a_n}$\} is the joint allocation decision by assigning $w_i$ to $h_n$.
    \item $\mathcal{T}$ := ${P}\left(s^{\prime} \mid s, a\right)$ is the state transition function.
    \item $\mathcal{R}$ := $f_R\left(s \times o, s^{\prime} \times o^{\prime}\right)$ is the reward function, which gives immediate reward after the transition from $s \times o$ to $s^{\prime} \times o^{\prime}$ by taking action $a$.
    \item $\gamma$ is the discount factor.
\end{itemize}

This formulation allows fundamental modeling of environment dynamics of the affective workload allocation problem in MH-MR teams, and explicitly defines various attributes of the team, such as the human state, including subjective and objective CWLs, and contextual information, including performance metrics, current workload distribution, and other system characteristics. The goal of this POMDP is to find the optimal policy $\pi^{*}: (s \times o) \mapsto {a}$ that obtains maximum system performance, i.e., the expected total discounted reward $\mathbb{E}[\sum_{k=0}^{\infty} \gamma^{k} r_{t+k}]$. Through this optimization process, the AWAC algorithm learns to adaptively adjust workload distribution for $n$ human agents based on contextual information and human conditions, including subjective and objective measurements of CWLs, to maximize operational performance.

\subsection{Data-driven Human Performance Model (HPM)}
\label{human_performance}
While the introduction of the POMDP and DRL can provide a better chance to find the optimal workload allocation strategy, it requires a high volume of interaction rounds to reach good performance. Therefore, a DRL model is typically trained in a simulation environment to achieve results in a cost-effective and timely manner. To build a sound simulation environment for an MH-MR team, the key challenge is to build a human performance model (HPM). that can simulate human performance based on the human state and serve as the transition function of the human state. 
To address this challenge, we propose a data-driven HPM that estimates the human operator's current mission performance from the subjective and objective measurements of the CWLs. This HPM can be easily adapted to various applications by tuning the parameters of the equations derived from the empirical experiments.

For the generalized HPM, we applied the Yerkes-Dodson law \cite{yerkes1908relation}, which is a psychological principle that describes the relationship between CWLs and mission performance. It states that performance generally improves with increased CWLs, but only up to a certain point. Beyond this point, further increases in CWLs lead to a decline in performance. The law is represented by an inverted-U shape as illustrated in Fig. \ref*{fig:human_yerkes_dodson} on Appendix \ref*{apx:figures}, and can be mathematically described by the Gaussian distribution as Eq. \ref{eq:ch05_performance_isa_pred}, with the optimal level of CWLs for maximum performance at the peak of the curve. The law helps explain why too much \updated{CWLs} can have a negative impact on performance and why finding the right level of \updated{CWLs} is important for optimal performance. The Yerkes-Dodson law has been applied to various fields to estimate human's performance in HRI research \cite{sam2021investigating}. 
We also found a similar relationship between the operator's CWLs and mission performance in our previous study \cite{jo2024mocas}.

\begin{equation} 
    \label{eq:ch05_performance_isa_pred}
        P(\mathcal{S}) = \frac{A}{{\sigma \sqrt {2\pi } }}e^{{{ - \left({\mathcal{S} - \mu } \right)^2 } \mathord{\left/ {\vphantom {{ - \left( {\mathcal{S} - \mu } \right)^2 } {2\sigma ^2 }}} \right. \kern-\nulldelimiterspace} {2\sigma ^2 }}} \\
\end{equation}

\noindent where area $A$ is calculated using the trapezoidal rule: $A = \sum_{i=0}^{n-1} (x_{i+1} - x_{i})(y_{i+1} - y_{i})/2$. The maximum and minimum values of ${S^s}$, ${S^o}$, and $P$ are used in the calculation. The values of $\sigma$ and $\mu$ are determined by the type of mission and the measurement methods used to evaluate human performance, such as subjective measurement, $P_{S^s}$ and objective measurement, $P_{S^o}$. $\mathcal{S}$ is the measured CWL and serves as an input variable to convert into $P$.

Then, the integrated human operator's mission performance model using $P_{S^s}$ and $P_{S^o}$ is estimated by 
\begin{equation}
    \label{eq:ch05_all_performance}
    \begin{split}
    P_{h_n} = \alpha_{p}P_{S^o}  + \beta_{p}P_{S^s}.
    \end{split}
\end{equation}
Here, we added two weights ($\alpha_{p} = 0.5$, $\beta_{p} = 0.5$) for the sensitivity of the proposed controller and type of the mission. We utilized both $P_{S^s}$ and $P_{S^o}$, which are calculated by subjective questionnaires and objective measurements of the CWL, respectively, in order to protect against unexpected malfunctions of the physiological sensors as the objective measurement and fake answers of the subjective questionnaires made by human operators intentionally. 

In order to allocate the workloads based on the estimated operator's mission performance $P$, we developed the DRL-based AWAC to find optimal changes for multi-human operators $\mathcal{H}, H \in \{h_1, \dots, h_n\}$, by comparing the current team mission performance with the predicted next team mission performance that reflects changes in the two variables of $\mathcal{W}$, which are the changed workloads of human agents:

\begin{equation}
    \label{eq:ch05_predicted_ISA_preidction}
    \begin{split}
    {S^{o}}_{t+1} = S^{o}_{t} + \Delta w
    \\
   {S^{s}}_{t+1} = S^{s}_{t} + \Delta w
    \end{split}
\end{equation}
where $\Delta w$ is the variance of the assigned workloads. The next performance is estimated using Eqs. \ref{eq:ch05_performance_isa_pred} and \ref{eq:ch05_all_performance}, as
\begin{equation}
    \label{eq:ch05_predicted_performance}
    \begin{split}
       P(S_{t+1}) = \alpha_{p}P({S^{o}}_{t+1}) + \beta_{p}P({S^{s}}_{t+1}).
    \end{split}
\end{equation}
These values are utilized in the AWAC algorithm to assign the optimal workloads based on each human operator's performance.

\subsection{Proximal Policy Optimization for Workload Allocation}
Based on the HPM, we built a DRL model to allocate optimal workloads for enhancing team performance. To train our DRL model, we assume that the objective measurement of the CWL predicted by APM indicates the human operator's CWL, and there are no fake answers on the subjective measurement of the CWL. Fig. \ref{fig:drl_diagram_model} depicts the learning diagram used in the proposed DRL model to find optimal transitions of the workloads based human operators' CWLs.

The state space ($\mathcal{S}$) in the DRL model is designed to consider individual CWLs measured by the self-reporting questionnaire ($S^{s}_{h_n}$) and by the predicted CWL by APM ($S^{o}_{h_n}$), where $n$ is the number of the human operators and $n \in \{1, \dots , n\}$. 
$S^{o}$ represents the objective CWL, and $S^{s}$ represents the subjective CWL.
The action space ($\mathcal{A}$) represents the assigned workloads based on operators' performance, $A \in (a_{1}, \dots, a_{n})$, which are estimated based on operators' $S^{o}$ and $S^{s}$.
We designed the team mission performance-based reward ($\mathcal{R}$) in order for the DRL model to achieve high team mission performance by comparing predicted performance after taking the next state ($\mathcal{S}^{\prime}$) and action ($\mathcal{A}^{\prime}$) with the current state and action. 

We built an environment to train our DRL model, utilizing the predefined state space $\mathcal{S}$, action space $\mathcal{A}$, and reward $\mathcal{R}$, through the use of OpenAI gym (see Algorithm \ref*{alg:drl_env_gym} on Appendix \ref*{apx:algorithm}). This was done in conjunction with our surveillance environment, as depicted in Fig. \ref{fig:drl_diagram_model}. Using PPO, we trained the DRL model on the environment to obtain the optimal policy $\pi$\cite{schulman2017proximal}. This process is terminated if $P_{team}^{\prime}$ is less than $P_{team}$, or if the sum of the assigned $\mathcal{A}$ falls outside the range of $\mathcal{W}$. A total of 1,000,000 samples were used to train the DRL model using three episodes, covering various workload allocation methods.

To validate the performance of our AWAC model, we compared it with a random workload allocation method. Both models were provided with equivalent inputs ($\mathcal{S}$) and subjected to equivalent restrictions as used in the HPM, along with the sum of workloads. We then measured the performance of each model based on the inputs, repeated the experiment 10,000 times, and performed repeated measures ANOVA (rmANOVA) tests. Our results indicate that the proposed model outperforms the random workload allocation method ($p < .01$).

\section{Case Study in CCTV Surveillance Scenarios}
\label{sec:design_user_exp}
In this section, we describe how the proposed AWAC can be applied to real MH-MR CCTV surveillance scenarios and how the AWAC, with various allocation methods, impacts team performance through exploratory user experiments. Moreover, this section explains the details of the CCTV monitoring task, the design of the user experiment, and the overall system configuration used in the user experiment.

\subsection{Team-based CCTV Monitoring Task}
To validate the proposed AWAC method, we designed a team-based user experiment that involved the MH-MR CCTV monitoring task, which required multiple human operators to monitor and control multiple agents/robots simultaneously. Such surveillance missions are widely required in diverse MH-MR system task scenarios, including security monitoring \cite{wang2008affective}, air traffic management \cite{pacaux2002common}, and performance checking \cite{stowers2017framework}. The surveillance scenario is typical in real-world human-agent systems, in which human operators undertake a simultaneous CCTV monitoring task while multiple sensors track them.

\begin{figure}[t]
    \centering
    \begin{subfigure}[b]{1\linewidth}
        \centering 
        \includegraphics[width=1\linewidth]{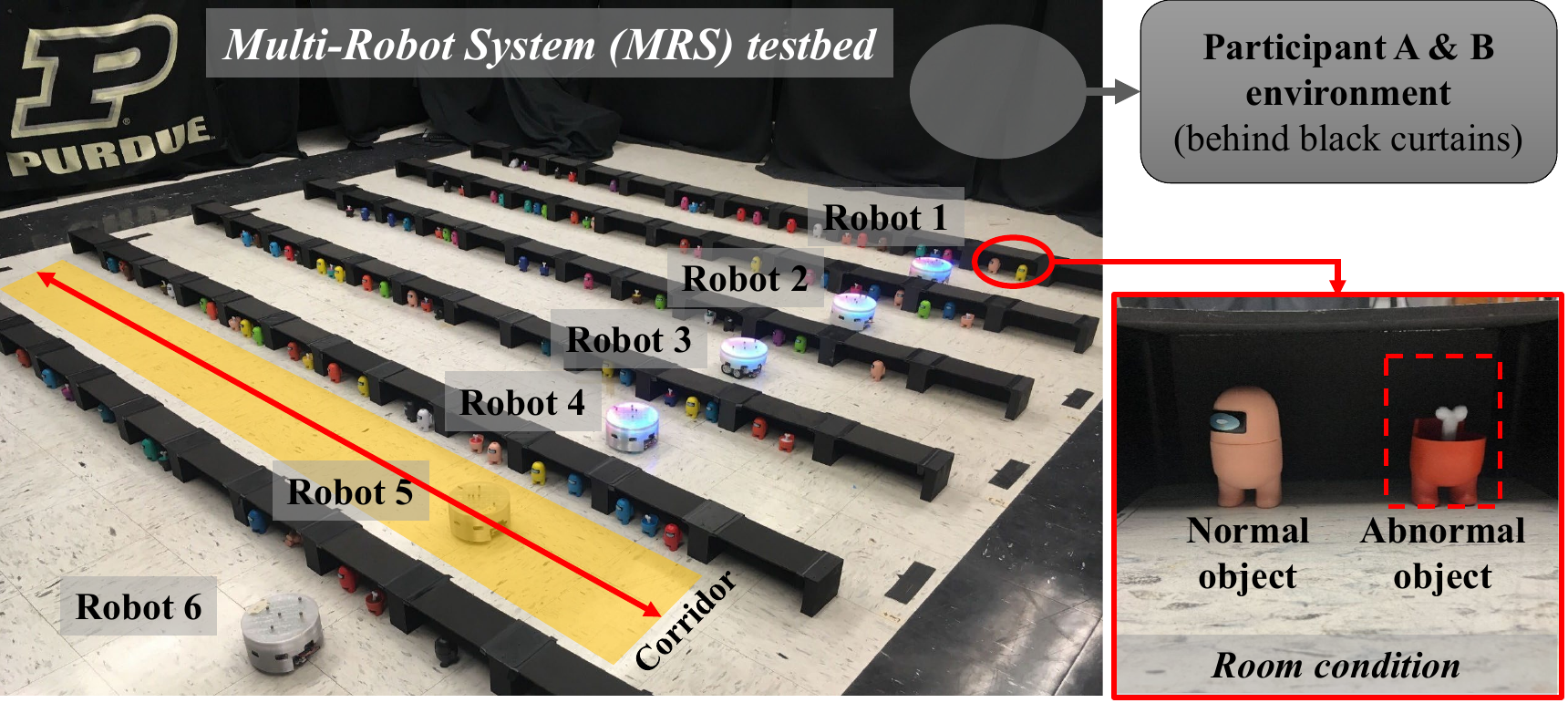}
        \caption{}
        \label{fig:mrs_testbed}
    \end{subfigure} 
   
    \begin{subfigure}[b]{1\linewidth}
        \centering 
        \includegraphics[width=1\linewidth]{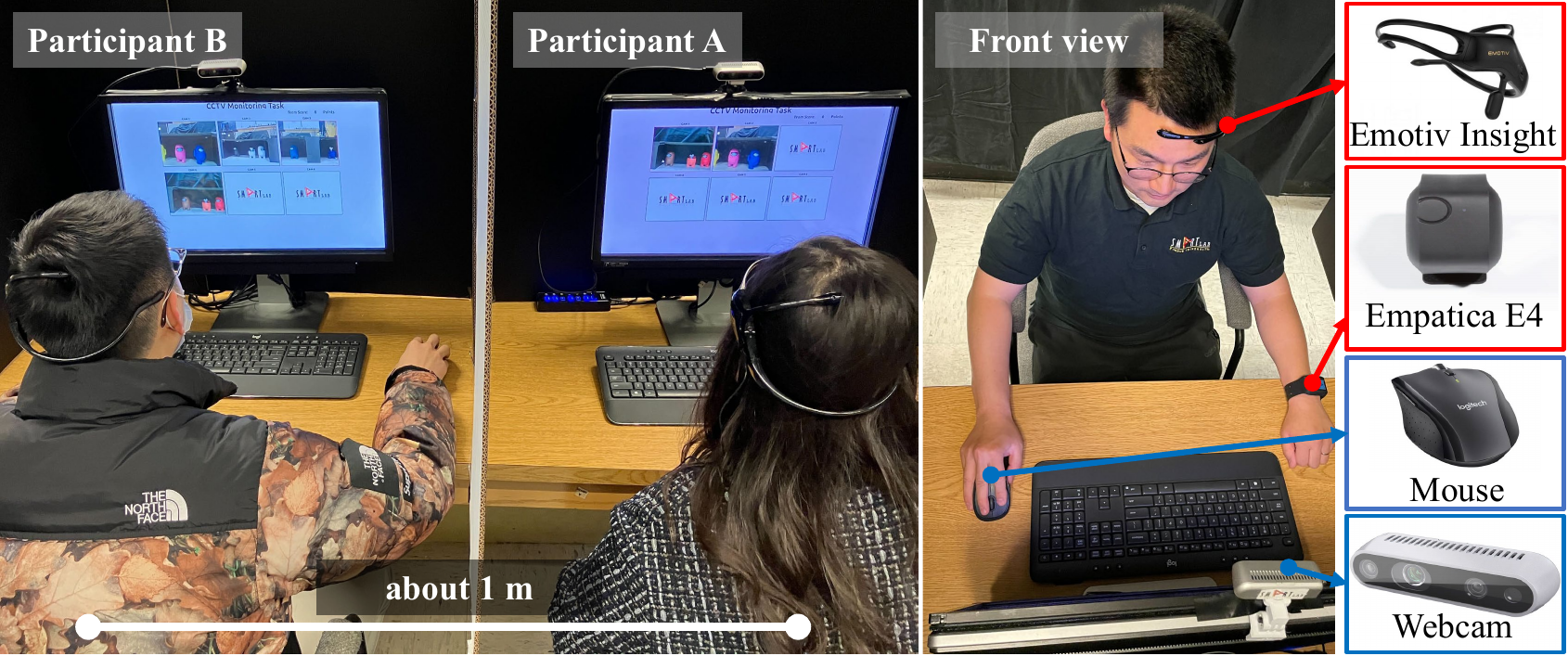}
        \caption{}
        \label{fig:phase_2_environment}
    \end{subfigure}
    
    \caption{Illustration of (a) MRS testbed for conducting surveillance missions and (b) Participants' workspace and wearable biosensors (red) and behavioral-monitoring devices (blue) used to collect physiological features and behavioral data.}
    \label{fig:experiment_setting}
\end{figure}

Based on the surveillance scenario, we built a generalized surveillance environment and team-based user study that supported multi-human operators in conducting CCTV monitoring tasks with multi-robot systems as MH-MR teams as illustrated in Fig. \ref{fig:experiment_setting}. Two human subjects work together as a team to conduct a simulated CCTV monitoring task together in an environment as shown in Fig. \ref{fig:phase_2_environment}. Both human subjects have identical roles and responsibilities in the team. During the CCTV monitoring task, the MRS platform moved through the corridor (shown in red in Fig. \ref{fig:mrs_testbed}) at a speed of approximately 300 $mm/s$ and streamed the room conditions to the participants' screens to simulate a CCTV monitoring task, as the participant's monitor screen in Fig. \ref{fig:phase_2_environment}.
The CCTV monitoring task is to click on the window displaying a streaming camera view containing an abnormal object, as illustrated in the right bottom of Fig. \ref{fig:mrs_testbed}: abnormal objects (e.g., skeleton objects) and normal objects. The objects were randomly placed in separated rooms, as shown in the left of Fig. \ref{fig:mrs_testbed}, and differed in color and quantity. In the CCTV monitoring task, one operator can monitor up to five cameras. This maximum number of camera views was determined through our previous experiment \cite{jo2024mocas}.

\subsection{User Study Procedure}

\begin{figure}[t]
    \centering
    \includegraphics[width=0.95\linewidth]{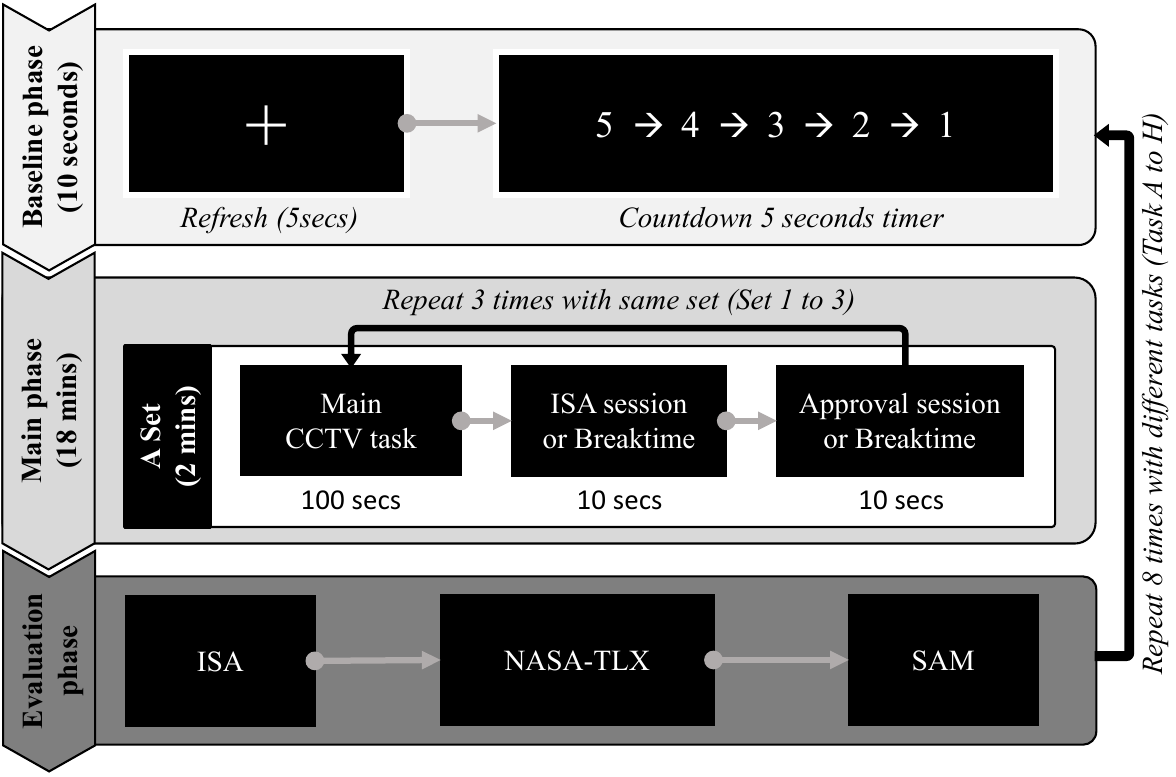}
    \caption{The experiment protocol used in the team-based user experiment involves three phases: Baseline, Main, and Evaluation phases. In the Main phase, a set is repeated three times for the CCTV monitoring task. This procedure is repeated eight times with different workload allocation methods randomly selected from Task A to H.}
    \label{fig:user_experiment}
\end{figure}

As illustrated in Fig. \ref{fig:user_experiment}, the participants conducted the eight tasks with random order (from Task A to H). This study was designed as a 2$\times$2$\times$2 within-subject study. This user experiment was reviewed and approved by the Purdue University Institutional Review Board (IRB) (\#IRB-2021-1813). 

For the team-based user experiment, we recruited 32 participants who met the health requirements (female: 9 and male: 23; see more details about participant requirements on Appendix \ref*{apx:part_req}) for 16 teams and conducted the experiments to validate the proposed AWAC, as shown in Fig. \ref*{fig:all_teams_study} on Appendix \ref*{apx:figures}. The participants had an age range of 18 to 34 (\textit{Mean}=23.81, \textit{S.D.}=4.17). Each participant was compensated \$15 for their time and efforts. To increase the subjects' engagement in the study, we provided additional compensation based on the overall team's scores on the given tasks.

Prior to starting the experiment, we explained the entire experimental procedure to two participants and asked them to fill out an informed consent and a demographic questionnaire. The participants were then instructed to wear two wearable biosensors. The wearable biosensor data were utilized to predict the participant's objective CWL. After completing the survey and sensor calibration process, we provided instructions for each of the nine tasks (one training and eight main experiment tasks). Each task consisted of three sets (total 300 seconds) and three break times (total 60 seconds), as illustrated in Fig. \ref{fig:user_experiment}. 
Then, participants performed a trial CCTV monitoring task using the CCTV monitoring graphic user interface (GUI) program as depicted in Fig. \ref*{fig:surveillance_GUIs} on Appendix \ref*{apx:figures}. The objective of the trial session was to familiarize themselves with the experiment hardware and to understand the CCTV monitoring tasks used in this experiment. Then, the participants conducted one of the tasks for 360 seconds and repeated it eight times under different experimental conditions. After completing each task, participants were asked to evaluate their emotional and cognitive states using questionnaires such as self-assessment manikin (SAM), instantaneous self-assessment (ISA), and NASA task load index (NASA-TLX) via the GUI programs.

At the end of the user experiment, participants were interviewed for approximately five minutes to gain insight into their overall experience with our MH-MR AWAC system. The interview began with a lead-off question (e.g., \textit{``Did you notice any differences between the eight sessions? If so, what were they?''}) and was followed by several other questions (e.g., \textit{ ``What are your thoughts on the ISA and Approval sessions during the mission?''}, and \textit{``Which method do you prefer for conducting this CCTV monitoring task: changing or fixed workloads?''}).

\subsection{Details of Tasks in the User Experiment}
In Task A, two participants view a fixed number of camera views (e.g., three camera views). In Task B, two participants discuss and decide on the allocation of workload (e.g., the number of camera views) before starting the main task, known as a consensus step, and view the fixed number of camera views based on the outcome of the discussion. In Task C and Task D, AWAC adjusts the workloads based on the subjective CWL reported by ISA, called ISA Session (IS), with an additional workload transition method known as Approval Session (AS) used to seek consent from the other participant before changing the workload. Task C has AS, while Task D does not. 
In Task E and Task F, AWAC automatically adjusts the number of camera views based on the objective CWL predicted by APM, called Prediction Session (PS). Task E has AS, while Task F does not. In Task G and Task H, AWAC automatically adjusts the number of camera views based on both subjective and objective measurements of the CWL (e.g., IS and PS). Task G has AS, while Task H does not. The tasks are summarized in Table \ref*{tab:ch5_task_list}. A supplementary video that demonstrates the user study experiment can be found at \url{https://youtu.be/qrmAQqfdLZk} and/or on Appendix \ref*{apx:video}.

\begin{table}[h!]
    \centering
    \caption{Summary of tasks used in the user experiment.}
    \label{tab:ch5_task_list}
    \resizebox{0.9\columnwidth}{!}{%
    \begin{tabular}{|c|c|c|c|c|}
        \hline
        \rowcolor[HTML]{C0C0C0} 
        \textbf{Task} & \textbf{\begin{tabular}[c]{@{}c@{}}Fixed \\ workload\end{tabular}} & \textbf{\begin{tabular}[c]{@{}c@{}}ISA \\ session (IS) \end{tabular}} & \textbf{\begin{tabular}[c]{@{}c@{}}Prediction \\ session  (PS) \end{tabular}} & \textbf{\begin{tabular}[c]{@{}c@{}}Approval \\ session (AS) \end{tabular}} \\ \hline\hline
        \textbf{A} & O & X & X & X \\ \hline
        \textbf{B} & O & X & X & O \\ \hline
        \textbf{C} & X & O & X & O \\ \hline
        \textbf{D} & X & O & X & X \\ \hline
        \textbf{E} & X & X & O & O \\ \hline
        \textbf{F} & X & X & O & X \\ \hline
        \textbf{G} & X & O & O & O \\ \hline
        \textbf{H} & X & O & O & X \\ \hline 
    \end{tabular}
    }
\end{table}

\begin{figure}[t]
    \centering 
    \includegraphics[width=0.9\linewidth]{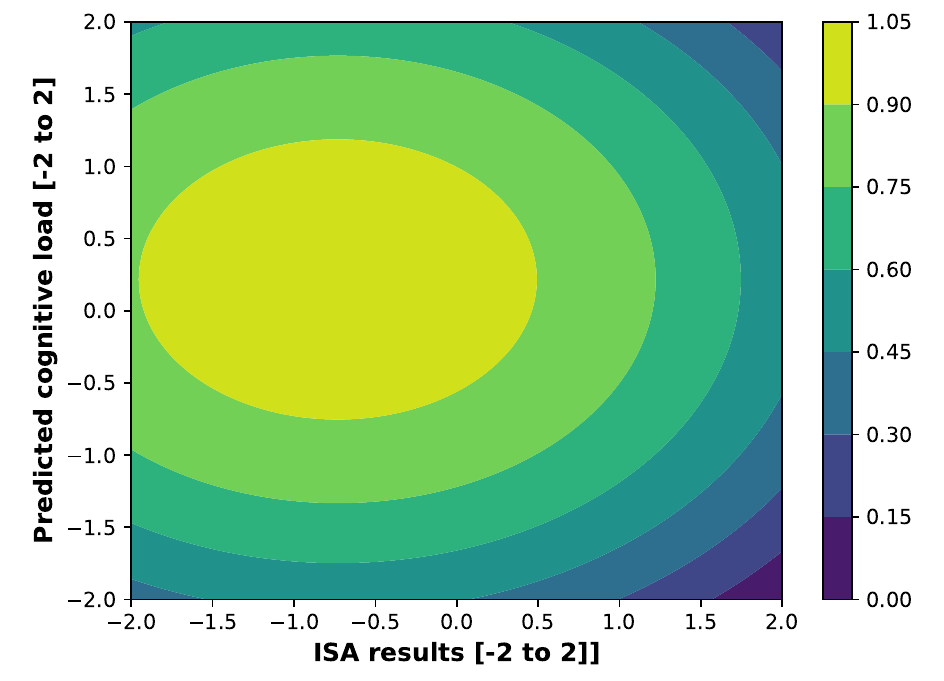}
    \caption{An illustration of the HPM calculated using the ISA score and predicted cognitive workload.}
    \label{fig:human_performance_model}
\end{figure}

\begin{figure*}[t]
    \centering
    \includegraphics[width=1\linewidth]{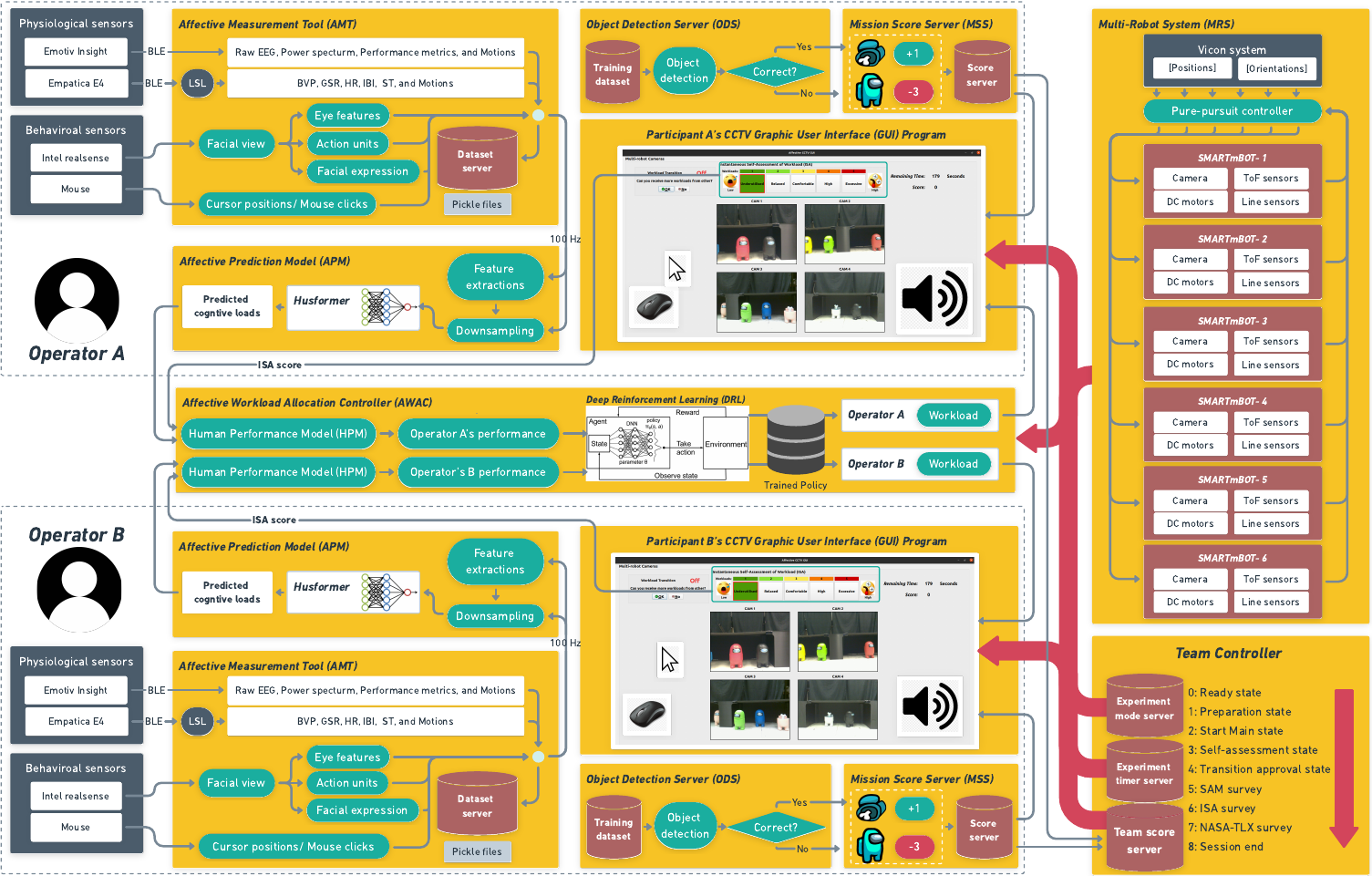}
    \caption{System architecture for the MH-MR surveillance task. More details can be found at the supplementary website: \url{https://sites.google.com/view/affective-workload-allocation}.}
    \label{fig:surveillance_system_mhmr}
\end{figure*}

\subsection{Tuning HPM parameters}
Using multimodal dataset for objective cognitive workload assessment on simultaneous tasks, called MOCAS \cite{jo2024mocas}, we analyzed the results of the subjective and objective measurements of CWL, obtained from ISA answers and APM prediction results.
Then, we tuned the parameters of Eq. \ref{eq:ch05_performance_isa_pred} based on the empirical results of the extensive MOCAS dataset. The tuned HPM can estimate the current CCTV monitoring task performance of the human operator from the subjective and objective measurements of CWL. The correlation between performance and subjective and objective CWL is illustrated in Fig. \ref{fig:human_performance_model}.

Each surveillance mission with MH-MR teams consists of three sets. To achieve better performance in each episode, we defined the reward function as $r = 0.33$ if the next team mission performance ($P_{team_{t+1}}$) is greater than the current team performance ($P_{team_{t}}$), and 0 otherwise. The $P_{team}$ is the mean of all human operator's mission performance, and $P_{team^{\prime}}$ is the predicted team mission performance on the next step, calculated by reflecting transitions in the number of camera views. We assume that as the number of camera views ($w_{i}$) increases, the operator's CWL ($h_{n}$) and other state variables ($S^{o}, S^{s}$) will also increase in the next step. The correlation between the number of camera views and state variables was found through empirical results from \cite{jo2024mocas,jo2024smart}.


\subsection{Overall System Configuration}
We developed a team-based surveillance system involving two human operators and six multi-robot platforms, as shown in Fig. \ref{fig:surveillance_system_mhmr}. The hardware configuration used for the operators in this user experiment is identical and is mainly comprised of the APM, which serves as the main interface for reading physiological signals and behavioral features, and measuring the objective measurement of the CWL. The subjective and objective measurements of CWL are then utilized as inputs for AWAC to reallocate workload. In addition to the main interfaces, additional sub-interfaces were used to conduct the team-based surveillance mission during the user experiment, as outlined below.

\subsubsection{Affective Measurement Tool (AMT)}\label{monitoring_framework}

As illustrated in Fig. \ref{fig:surveillance_system_mhmr}, we developed the AMT to measure physiological and behavioral data from wearable biosensors and behavior-monitoring devices, respectively. We utilize a robot operating system 2 (ROS2)-based wearable biosensors package \cite{jo2020ros} to record human data over wearable biosensors for communicating with multi-robot systems supporting ROS2. This package allows for communication between wearable biosensors and multi-robot systems supporting ROS2. To provide real-time computing, support for multiple robots, and reliable streaming data from various nodes, each node in the biosensor package follows a standardized node and topic structure.

In the AMT, raw physiological signals from wearable biosensors are collected at a $100~Hz$ sampling rate, while behavioral features from the facial view are extracted at a $30~Hz$ sampling rate. This allows for efficient and accurate measurement of affective states. For the physiological signals, we used two off-the-shelf wearable biosensors, the Empactica E4 and Emotiv Insight, to collect physiological data, and a webcam (e.g., Intel RealSense) to record behavioral data. The Emotiv Insight provides readings of 5-channel EEGs, power spectrum (theta, alpha, beta, and gamma), performance metrics, and motion data, while the Empatica E4 provides readings of blood volume pulse (BVP), galvanic skin response (GSR), heart rate (HR), inter-beat interval (IBI), skin temperature (ST), and motion data. These measurements allow us to accurately and reliably capture affective states during collaborative tasks with multi-robot systems. For the behavioral data, we extracted various features from the facial camera views, such as eye-aspect ratio, facial action units, and facial expressions. These behavioral measurements provide insight into operator affective states and can be used to complement the physiological data collected by the wearable biosensors.

The data collected from wearable biosensors and behavioral features from the facial camera views in the AMT are downsampled to a sampling rate of $100~Hz$. This processed data is then used as inputs to predict operator affective states in real-time. This process allows the team to effectively collaborate and perform tasks efficiently, despite variations in operator affective states. By leveraging the high temporal resolution of these measurements, the AMT and APM can provide a detailed understanding of operator affective states and support the dynamic allocation of workloads in real-time.

\subsubsection{Affective Prediction Model (APM)}\label{dl_prediction}
As illustrated in Fig. \ref{fig:surveillance_system_mhmr}, we also use APM to predict CWLs from the objective measurement of physiological and behavioral signals, which can range from low to medium to high. We adopt the \textit{Husformer} \cite{wang2022husformer}, an end-to-end multimodal transformer framework for the recognition of multimodal human cognitive load, for building the APM. To make predictions of objective CWLs of operators, the APM uses the multimodal bio-signals collected through AMT as the input, and predicts the objective cognitive load levels, i.e., low, medium, and high, as the outputs at $100~Hz$. We refer readers to \cite{wang2022husformer} for more details of the APM.

\subsubsection{CCTV Monitoring GUI Program}
The CCTV GUI Program is the direct interface between the participant and the multi-robot system for conducting the surveillance task, as shown in Fig. \ref*{fig:surveillance_GUIs} on Appendix \ref*{apx:figures}. The CCTV GUI program displays multiple windows of the camera views and team scores while performing the task. The GUI starts with the setup screen to test communication among the wearable biosensors, behavioral monitoring devices, and multi-robot systems through ROS2. If there is no issue, the GUI enters the preparation step, showing a black cross for 5 seconds, followed by a 5-second countdown timer. Then, the GUI enters the main experiment step where the participant monitors multiple camera views simultaneously, called the CCTV monitoring task, to find abnormal objects on the screens. Additionally, only the team scores are displayed on the GUI to reduce peer and time pressures. 

Each task consists of three sets, and each set takes 100 seconds, with a break time of about 20 seconds between each set. The break time is mainly used for reporting participant's subjective CWL through the IS for 10 seconds, and the other 10 seconds are for AS. However, participants have break time if the current task does not require collecting both values (such as Task A and B). After each task, participants are redirected to three surveys, including SAM, ISA, and NASA-TLX, where they respond based on their feelings and thoughts about the entire task. We then repeated eight tasks for 90 minutes.

\subsubsection{Object Detection Server (ODS)}
It performs the CCTV monitoring task of detecting abnormal objects on the streaming video from the multi-robots displayed on the GUI program. The ODS checks whether the human operator detects abnormal or normal objects and provides audio feedback to the human operator based on the results of the ODS. 
If human operators detect an abnormal object, they are awarded 1 point (e.g., reward). However, if they detect a normal object, they lose 3 points (e.g., penalty). Based on our pilot test, we decided on two points: a higher penalty point (-3) than the reward (+1), to encourage careful consideration by human operators conducting CCTV monitoring tasks.
For detecting abnormal objects, we used the object detection algorithm, which has an accuracy of 99.96\% to detect abnormal objects among two classes (i.e., normal and abnormal objects).

\subsubsection{Mission Score Server (MSS)}
It manages the rewards (+1 point) and penalty (-3 points) based on the participant's performance in the CCTV monitoring task. The MSS works in conjunction with the ODS and GUI program during task execution. The ODS updates the MSS by determining if the human operator has completed the task correctly, while the GUI program serves as the user's direct interface and displays the updated team score as the task is being carried out.

\subsubsection{Multi-Robot System (MRS)}
The MRS consists of six ROS2-based multi-robot platforms, known as SMARTmBOT \cite{jo2022smartmbot}. 
The MRS consists of six ROS2-based multi-robot platforms, which is an open-source mobile robot platform. The Vicon motion capture system tracks the robot's locations, which uses reflective markers attached to the top of the robots. To perform the surveillance mission, a pure-pursuit control algorithm was employed, allowing the robot to repetitively travel between the start and goal position \cite{paden2016survey}.

\section{Results and Analysis}
\label{sec:result_analysis}  
In this section, we present the findings from our exploratory research. The user experiment is a team-based user study involving two human operators and MRS. The main objective of this experiment is to investigate the effect of the proposed AWAC on team performance and to find optimal workload allocation strategies by comparing with various task allocation strategies. Therefore, we adjusted a significance level of $\alpha$ to $.10$ for statistical analysis, which can provide valuable insights for future research regarding MH-MR teams.
Table \ref*{tab:standized_results_a} and Table \ref*{tab:all_anova_restuls_b} on Appendix \ref*{apx:tables} show the normalized team performance (or obtained mission scores) of all the teams. We normalized the raw team performance ($P$) by dividing it by the mean of the team performance ($\mu$); $P_{norm} = P/\mu$, in order to create standards and transform data taken from different teams into a consistent format.

\subsection{AWAC Validation and Results} 

\begin{figure}[t]
    \centering
    \begin{subfigure}{1\linewidth}
        \centering
        \includegraphics[width=1\linewidth]{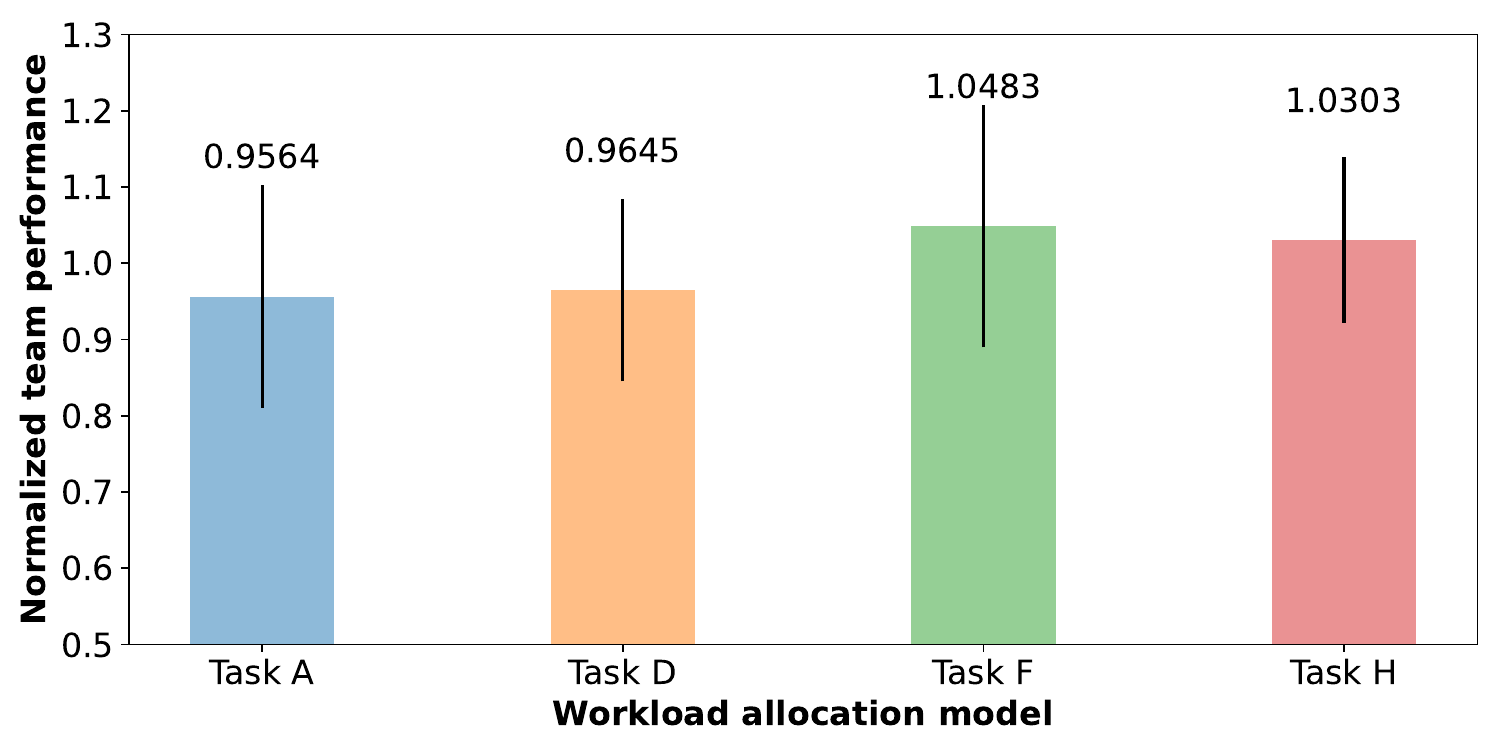}
        \caption{}
        \label{fig:drl_validation_exp}
    \end{subfigure}  
    
    \begin{subfigure}{1\linewidth}
        \centering
        \includegraphics[width=1\linewidth]{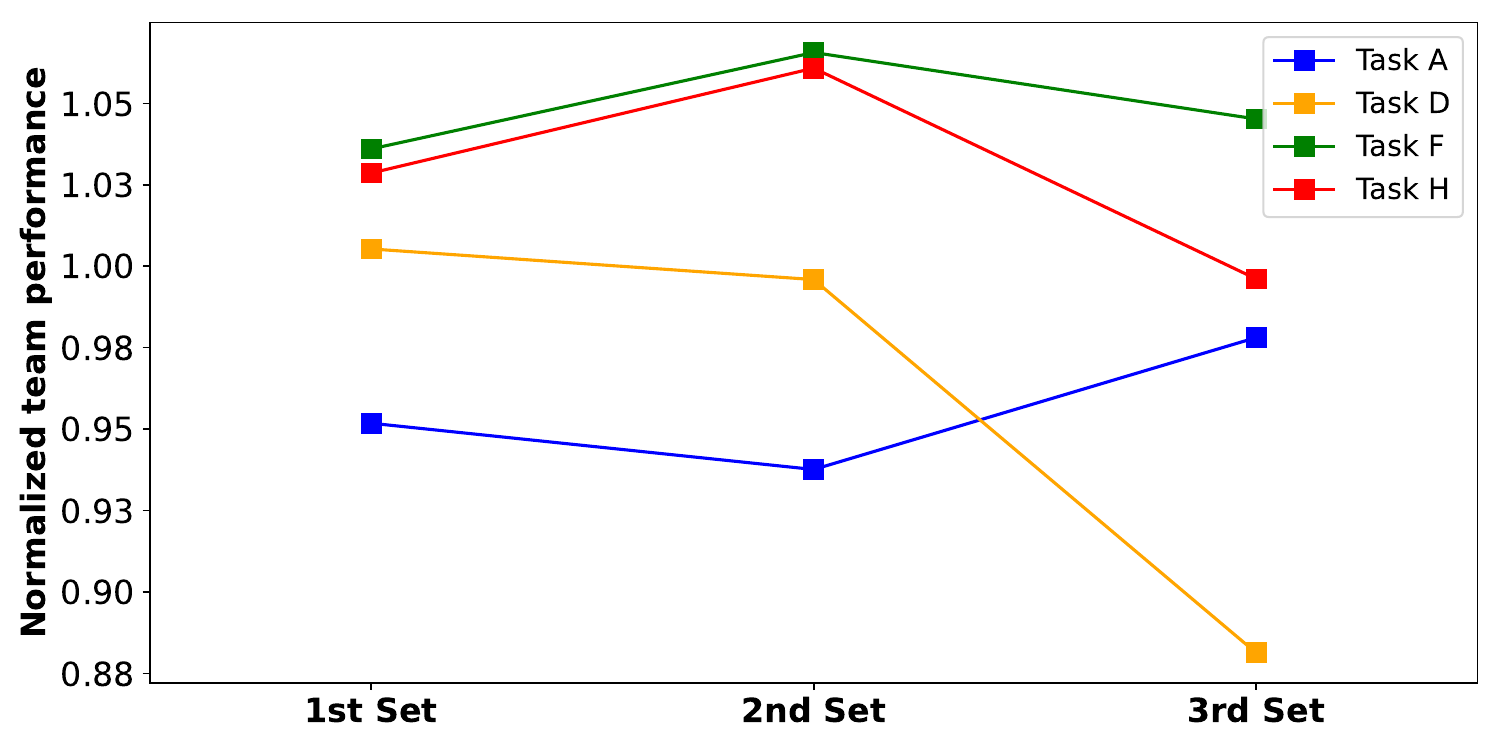}
        \caption{}
        \label{fig:phase_score} 
    \end{subfigure}
    \caption{Results of the comparison experiment to validate the AWAC on four tasks (Task A, D, F, and H): (a) Distribution of team performance on the four tasks used to validate the AWAC performance, and (b) Mean team performance obtained in each set (1st set, 2nd set, and 3rd set) of the four tasks.}
    \label{fig:drl_validation}
\end{figure}

In order to find the optimal combination for AWAC (e.g., fixed, PS, IS, and AS), we divided teams into two groups: Tasks A, D, F, H, and Tasks B, C, E, G, based on the presence of the AS. Then, we performed the rmANOVA test to investigate the effects of the proposed AWAC without AS on team performance. 
We then conducted the post-hoc analysis using a paired parametric t-test with Bonferroni correction to explore the contrasts among different workload allocation models \cite{kirk1974experimental}. The team performance across Task A, D, F, and H was compared using mean scores and standard deviations. Fig. \ref{fig:drl_validation_exp} illustrates the team performance of Task A, D, F, and H. 

The rmANOVA analysis revealed significant effects in the dependent variable among the groups ($F(3, 45)$ = 2.2134, $p$ = .0995, $\eta_{p}^{2}$ = .0813) with mean scores of 0.9564 for Task A, 0.9645 for Task D, 1.0483 for Task F, and 1.0303 for Task H (Table \ref*{tab:all_anova_restuls_a} on Appendix \ref*{apx:tables}). 
From the results of the post-hoc tests, we observed that Task F demonstrated higher performance than Task A ($T(15)$ = -1.8182, $p$ = .0891), although there are significant effects. Furthermore, Task H demonstrated higher performance than Task D ($T(15)$ = -2.1536, $p$ = .0479). However, no significant differences were found between Task A and D ($T(15)$ = -0.1713, $p$ = .8663), Task D and F ($T(15)$ = -1.6795, $p$ = .1138), and Task F and H ($T(15)$ = 0.4053, $p$ = .6909). These results underscore task-specific variations in team performance, with Task F and H generally demonstrating better performance compared to Task A and D, although no significant difference emerged between Task F and H.

Based on the rmANOVA result, we conclude that Task F and H, which utilized the proposed AWAC, produced higher team performance scores than the baseline task (Task A). The mean mission score in Task A was 372.25, but the other three tasks using AWAC were 376.375, 403.875, and 402.125, respectively. 
Therefore, we confirmed the proposed AWAC's effectiveness in improving team performance by adaptively allocating the operator's workload based on their affective state.
Given the similar team performance of Task A and Task D, we can assume that IS does not significantly affect the team performance compared to PS used in Task F and H.

In order to more deeply investigate the performance of our AWAC, we analyzed the team performance obtained in each set of the task; each task in our user experiment has three sets as illustrated in Fig. \ref{fig:user_experiment}. We calculated the team performance obtained in the 1st, 2nd, and 3rd sets, as illustrated in Fig. \ref{fig:phase_score}. We observed that the team performance obtained in each set of tasks with our AWAC (Task D, F, and H) was higher than that of Task A without AWAC in the 2nd set of the missions ($F(3, 45)$ = 2.5976, $p$ = .0639, $\eta_{p}^{2}$ = .0931). However, the team performance of Task D, F, and H decreased in the next 3rd set compared to Task A, but there was no statistically significant difference, so we could not say whether the team performance increased or decreased compared to the previous set.

\begin{figure}[t!]
    \centering
     \begin{subfigure}[b]{1\linewidth}
        \centering 
        \includegraphics[width=1\linewidth]{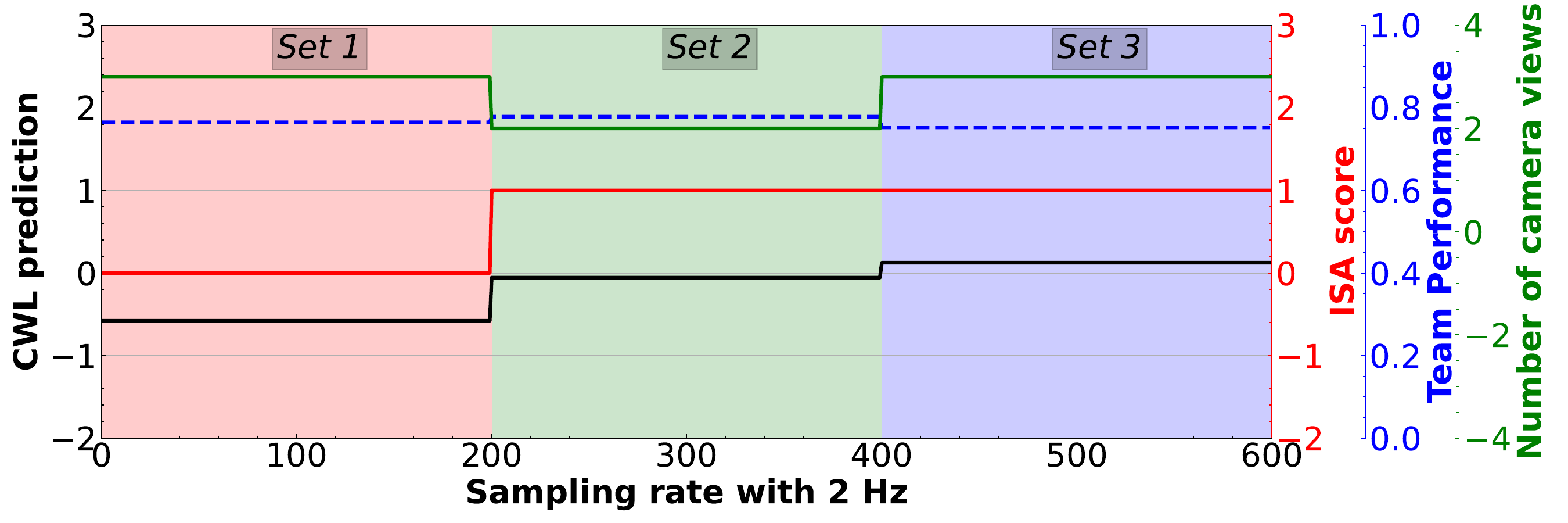}
        \caption{}
    \end{subfigure}
    
    \begin{subfigure}[b]{1\linewidth}
        \centering 
        \includegraphics[width=1\linewidth]{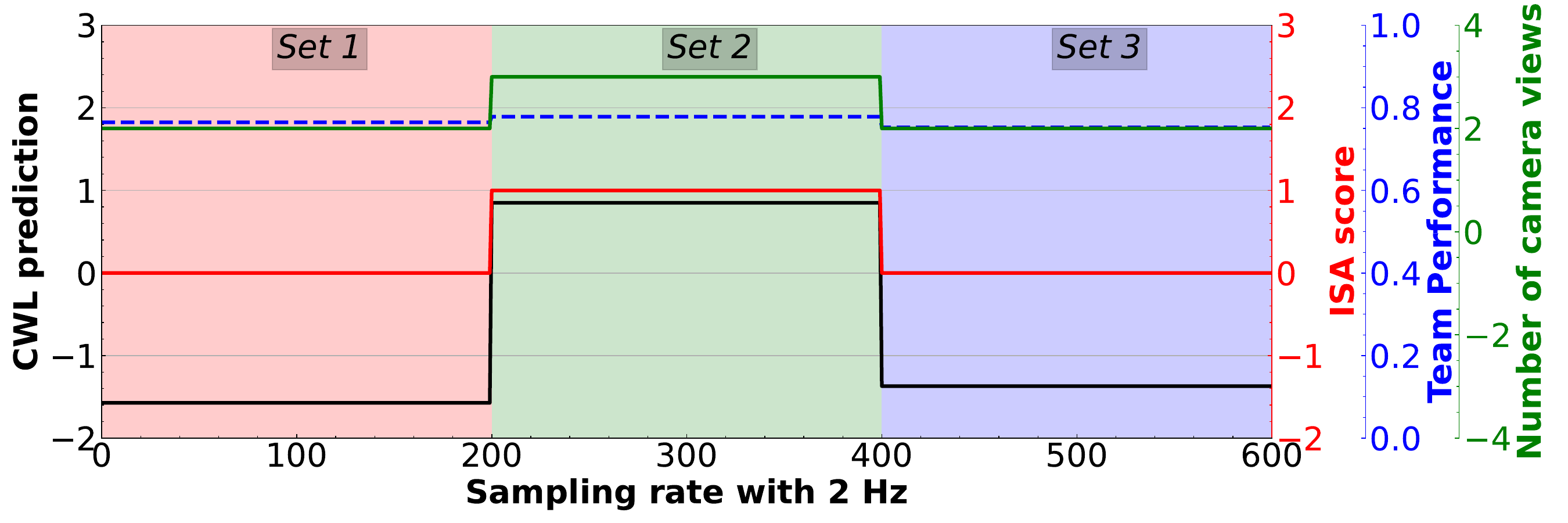}
        \caption{}
    \end{subfigure}
    
    \caption{Examples of workload allocation transitions generated by AWAC during user experiments. (a) Operator A's data and (b) Operator B's data. Red indicates ISA scores, blue indicates team performance, green indicates the allocated camera number, and black indicates predicted cognitive workload.}
    \label{fig:p_test_ps} 
\end{figure}

Thus, we can conclude that our AWAC plays a significant role in improving team performance. Additionally, we observed that the team performance of Task D suddenly dropped, which may imply that the subjective CWL responses were inaccurate, resulting in our AWAC reallocating the wrong workloads to each operator. We also noticed that the performance of Task A increased in the transition from the 2nd set to the 3rd set, but the team performance remained lower than tasks with AWAC. This observation suggests that our AWAC can effectively maximize operator performance in a shorter time than other allocation methods.

The subjective analysis of the effectiveness of the AWAC also supports the results of the objective analysis. 
As illustrated in Fig. \ref*{fig:sam_score} on Appendix \ref*{apx:figures}, we observed that the valence scores of the SAM questionnaire in the task with our AWAC are positive. This means that our AWAC can positively influence human operator's emotional states by inducing positive valence (e.g., happy), thereby improving the productivity and effectiveness of the missions. This is consistent with the Yerke-Dodson law \cite{yerkes1908relation} and other research findings \cite{du2020examining}. 

Fig. \ref{fig:p_test_ps} illustrates an example of workload reallocation recorded during the user experiment while participants were performing the CCTV monitoring task. The figure demonstrates that our AWAC successfully reallocated the workload for each participant based on their performance estimated from ISA scores and predicted CWL.

\subsection{Comparison of Workload Allocation Methods} 
Beyond validating the performance and effectiveness of the AWAC, we investigated the effects of various workload allocation methods on team performance for MH-MR teams. We designed three sessions as mentioned in Table \ref{tab:ch5_task_list}; IS, PS, and AS. The IS session measured the subjective CWL of human operators using a five-point rating scale during the missions. The PS session predicted the CWL of human operators using DL-based APM. The AS session asked the human operator to accept or reject the proposed workload transition. If the human operator rejected the workload change, the workload would not be altered.

To investigate the impact of the AWAC on team performance, we conducted the rmANOVA test with significance level of $\alpha$ to $.10$ by comparing Task B, C, E, and G. We also conducted post-hoc tests using a paired parametric t-test with Bonferroni correction. Fig. \ref*{fig:workload_method_exp} on Appendix \ref*{apx:figures} presents the team performance of Task B, C, E, and G. The rmANOVA test showed significant effects: in the dependent variable among the groups ($F(3, 45)$ = 2.3585, $p$ = .0842, $\eta_{p}^{2}$ = .0861) with Task B having a mean of 1.0275, Task C with 0.9316, Task E with 1.0046, and Task G with 1.037 (Table \ref*{tab:all_anova_restuls_b} on Appendix \ref*{apx:tables}).

Based on the results of the team performance analysis, it can be concluded that Task G, which utilized both IS and PS, can get higher team performance than other tasks (Task B, C, and E). It was also observed that the capability-based workload allocation (Task B) can get more team performance than tasks that only used IS or only used PS. Therefore, it can be suggested that workload allocation methods that include AS should consider both subjective and objective CWL (IS and PS) to optimize team performance.

In the tasks without AS (Task A, D, F, and H), we performed the rmANOVA test with significance level of $\alpha$ to $.10$ to compare the effects of the experimental conditions of the tasks on the team performance. Fig. \ref{fig:drl_validation_exp} shows the distribution of team performance on Task A, D, F, and H. The rmANOVA results showed there are significant effects on team performances based on the conditions ($F(3, 60)$ = 2.2139, $p$ = .0995, $\eta_{p}^{2}$ = .0813). The mean of Task A is 0.9564, the mean of Task D is 0.9645, Task F is 1.048, and Task H is 1.03. Thus, we can conclude that workload allocation methods without AS should consider applying the objective operator's CWL (e.g., PS) to maximize the team performance.

We performed the rmANOVA test to compare the effects of the experimental conditions of Task A and B on the team performance. Fig. \ref*{fig:comparison_a_b} on Appendix \ref*{apx:figures} shows the distribution of team performance on both tasks. The rmANOVA revealed no statistically significant difference between the two tasks ($F(1, 30)$ = 1.5959, $p$ = .2258, $\eta_{p}^{2}$ = .0640). Therefore, we can conclude that there is no difference between the task allocation based on the operator's capability (Task B) and the counterbalanced task allocation (Task A).

\subsection{Analysis of Subjective Questionnaires} 
After completing each task in the user experiment, participants were asked to rate their emotions and CWL using SAM, ISA, and NASA-TLX. We conducted rmANOVA tests with a significance level of $\alpha$ to $.10$ on the results of the subjective questionnaires and found no statistically significant difference between tasks in all questionnaires ($F(7, 255)$ = 1.7411, $p$ = .3474). However, we observed that Task G with PS and IS (\textit{Mean}=0.91, \textit{S.D.}=1.99) and Task E with PS (\textit{Mean}=0.91, \textit{S.D.}=2.07) had positive effects on the participant's emotions, especially valence, compared to the other tasks. The SAM questionnaire's valence values distribution is illustrated in Fig. \ref*{fig:sam_score} on Appendix \ref*{apx:figures}.

\subsection{Analysis of Post-Interview} 
After completing all tasks, we conducted interviews with the participants to gather feedback on our AWAC system. Most of the participants preferred changing the workload for the CCTV monitoring task rather than a fixed workload. Some participants reported that IS and AS were helpful in conducting missions (see the extended post-interview analysis in Appendix \ref*{apx:post_interview}).

From the post-interview sessions, we interestingly observed that some participants felt sorry for getting lower mission scores compared to their teammates, especially if they were friends.
In order to investigate the effect of the friendship between two operators on the team performance, we conducted a one-way ANOVA test to compare the performance of two groups (e.g., \textit{Group A} and \textit{Group B}) in from Task A, which is our baseline of the task allocation methods without any other control variables, such as IS, PS, and AS. \textit{Group A} consists of Teams 1, 7, 8, 9, and 10, those who know each other. \textit{Group B} consists of other teams, those who do not each other. According to the results of the one-way ANOVA test, there is no significant difference between Group A and B ($F(1, 15)$ = 0.0182, $p$ = .8945, $\eta_{p}^{2}$ = 0.036; See Fig. \ref*{fig:friendship_compare}). This indicates that there is no correlation between the friendship of two operators and the team's performance. Therefore, we can conclude that friendship between team members does not affect team performance.

\subsection{Summary of Findings}
We validated the performance of the AWAC and found there is a significant difference at $\alpha$ to $.10$ from our exploratory user experiment. The tasks with AWAC can achieve better team performance scores compared to tasks without AWAC, suggesting that reallocating the workload based on the operator's CWL for the team mission has positive effects on team performance. Among the four tasks (Task A, D, F, and H), the task with the PS achieved the best team performance, indicating that predicting the operator's CWL may play an important role in workload allocation for MH-MR teams.

When AS was provided, having both the IS (i.e., subjective measurement) and the PS (i.e., objective CWL measurement) achieved the best team performance. In addition, having the PS performed better than having the IS, suggesting that the objective measurement may be a better option for achieving better team performance. Furthermore, when the subjective and objective sessions were not provided, and only AS was given, the performance was better than when the subjective and objective measurements were provided separately. This suggests that workload allocation through consultation among team members can be more effective than workload allocation through subjective and objective measurements. However, as mentioned above, this consultation-based approach was not better than when all three sessions (AS, PS, and IS) were provided. In other words, the best performance in the CCTV monitoring task introduced in this study can be achieved by allocating the same workload in the beginning and soon after implementing the proposed AWAC with all three sessions.

\section{Discussion} 
\label{sec:discussion}
We introduced the DRL-based AWAC that enables human operators to perform better with teammates and multi-robot systems through team-based user experiments in a CCTV surveillance scenario involving MH-MR teams. The AWAC intelligently assigns appropriate workloads based on individual and team performance metrics, which are calculated using a combination of subjective CWL reported through self-reporting questionnaires and objective CWL predicted by our DL-based APM. Furthermore, we compared different workload allocation strategies for MH-MR teams and found that consulting with team members to allocate workload is more efficient than relying solely on our AWAC, which reflects both subjective and objective CWLs.

To evaluate the effectiveness of our AWAC, we conducted an rmANOVA with a significance level of $\alpha$ = $.10$ to analyze the mission scores obtained during the experiments. The results indicated that our proposed DRL-based AWAC led to better team performance. However, the significant effects were observed only at $\alpha$ = $.10$ and not at the conventional $\alpha$ = $.05$, as our study is exploratory, aimed at finding the optimal AWAC settings (Task A to Task H) and investigating the effects of the AWAC on team performance. The relatively small sample size in our team-based user experiment also contributed to these results. Although the statistical findings do not meet the $\alpha$ = $.05$, we observed significant effects of using affective states on team performance. Additionally, our post-analysis showed that the p-value decreased as more teams were included in the analysis (from Teams 1-4, 1-8, 1-12, to 1-16), suggesting that further research with a larger sample could yield more significant results ($p < .05$). Fig.~\ref{fig:p_treand} on Appendix~\ref*{apx:figures} illustrates the reduction of p-value with different numbers of teams.

In addition, generalization errors can occur in the proposed HPM and the DL-based APM. The HPM estimates human operators' mission performance using both self-reported and predicted CWL via the ISA and our DL prediction model, respectively.
To develop the HPM, we utilized the ISA scores and predicted CWLs of human subjects from the dataset we built in a previous study \cite{jo2024mocas}. From the dataset, we observed that the CWLs differ depending on the number of camera views while performing the surveillance task, and there are positive correlations between NASA-TLX and ISA scores ($\gamma > 0.5$). Therefore, we assumed that most participants had similar CWLs depending on the number of camera views. For instance, when performing our surveillance mission with one camera view, their CWL is low, and if they use more camera views, the CWL increases. We applied this knowledge to develop the HPM.

The DL-based APM was also developed using the same dataset collected from our previous experiment \cite{jo2024mocas}. However, due to participants' facial masks, our facial feature extraction programs failed to extract some of the facial features (such as eye aspect ratio and action units) from the subjects. Therefore, we only used 70\% of the MOCAS dataset for training the DL-based APM. To compensate for the small size of the training dataset, we applied \textit{K}-fold cross-validation ($K$=5), which is a re-sampling technique that generates more data from a limited data sample. Thus, we assumed that the prediction results could represent the current human operator's CWL while performing the CCTV monitoring task.

To ensure the sensibility of the proposed DRL-based APM, we added weights ($\alpha$ and $\beta$) to Eq. \ref{eq:ch05_all_performance}, which estimates human performance using both subjective and objective measurements of the CWL. During our team-based user experiment, we defined $\alpha_{p}$ and $\beta_{p}$ as 0.5 each, which were decided based on our pilot test. However, we may need to adjust these weights depending on the accuracy of the DL-based prediction algorithm. If the prediction results of the algorithm are reliable and high accuracy, we may need to increase the weight of $\beta_{p}$ to allocate the workload appropriately based on the human operator's performance. 

In this research, we focused only on the number of camera views as one of the primary factors (e.g., robot speed) that can directly affect an operator's CWL. Then, we utilized our HPM to convert the operator's CWLs into their performance, which determines the operator's workload allocation. This decision was made to focus on validating the effectiveness of monitoring human affective states on MH-MR team missions and finding the optimal workload allocation methods for MH-MR teams. However, we have identified another factor through previous experiments, which is that the robot's speed can also impact mission scores. In a previous study \cite{jo2024smart}, we observed that the mission scores of participants were influenced by the robot speed ($p < .001$).  

Moreover, the proposed AWAC was only validated through a user experiment involving two human operators and six multi-robot platforms due to limitations in experiment space and equipment, raising concerns about its scalability. However, this issue can be addressed by expanding our DRL framework to include more human operator actions ($\mathcal{A}$) and states ($\mathcal{S}$) or multi-robot platforms related to the total workloads. It is important to note that the number of human operators should be less than the number of robot agents to effectively allocate workloads based on operators' performance for the mission.

Based on the results of our experiment, our proposed workload allocation method (AWAC) that considers both subjective and objective CWL measurement (i.e., IS and PS) and incorporates the operator's opinion on workload transition (i.e., AS) led to better team performance compared to traditional workload allocation methods (Task A). The most common workload allocation method in our society is the capability-based workload allocation under the agreement, which allocates workloads based on the operator's preference, ability, skills, experience, and so on \cite{moacdieh2020effects,wigboldus2004capacity}. However, capability-based workload allocation methods are difficult to handle sudden changes in the operator's capability during the mission, so \cite{mina2020adaptive} proposed utilizing objective CWL measurement methods using physiological sensors (i.e., PS) to overcome this drawback. They found that CWL measurement-based workload allocation methods outperformed traditional allocation methods, which was also observed in our study. Specifically, objective-measurement-based workload allocation (Task E) outperformed capability-based workload allocation methods (Task C). Furthermore, we found that our proposed allocation methods using both IS and PS (Task G) outperformed traditional workload allocation methods (Task B and C) and objective CWL measurement-based workload allocation methods (Task E).

\section{Conclusion}
\label{sec:conclusion}
We introduced the deep reinforcement learning-based affective workload allocation controller (AWAC) that enables human operators to perform better with teammates and multi-robot systems. The AWAC intelligently assigns appropriate workloads based on individual and team performance metrics, which are calculated using a combination of subjective cognitive workload reported through the instantaneous self-assessment method and objective cognitive workload predicted by our DL-based affective physiological model. We evaluated the effectiveness of our AWAC and human performance model through team-based user experiments on a CCTV surveillance scenario involving multi-human and multi-robot teams. We used rmANOVA to analyze the mission scores obtained during the experiments, which showed that our proposed DRL-based AWAC resulted in better team performance. This indicates that workload allocation based on human operators' cognitive workload is critical to improving team performance. Furthermore, we compared different workload allocation strategies for MH-MR teams in the experiment, and found that allocating workload by consulting with team members is a more efficient method than relying solely on our AWAC that reflects subjective and objective CWLs. 

Our study highlighted the potential of the DRL-based AWAC in improving team performance in MH-MR team, but future research should focus on enhancing its robustness and applicability. Future work will involve conducting experiments with a larger and more diverse participant pool to improve the statistical power and generalizability of the findings, which will help validate the approach at a more significance level $\alpha$ to $.05$. Additionally, more detailed comparisons between specific tasks, particularly between Task A (fixed workload) and Task G (full AWAC), should be conducted to provide deeper insights into the effects of our AWAC over traditional methods. Exploring the scalability of the AWAC by testing it with larger MH-MR teams and more complex MH-MR setups is also planned. Finally, future studies should consider additional factors, such as robot speed, to further refine workload allocation and enhance overall team performance. Addressing these areas will help to validate the effectiveness of the AWAC more robustly and contribute to the development of more efficient workload allocation strategies in HRI scenarios.

\section*{Supplementary Materials}
This PDF file includes: Appendix \ref*{apx:video}. Supplementary Video \& Website; Appendix \ref*{apx:part_req}. Participant Requirement; Appendix \ref*{apx:algorithm}. Algorithm \ref*{alg:drl_env_gym}; 
Appendix \ref*{apx:post_interview}. Extended Post-interview Analysis; 
Appendix \ref*{apx:figures}. Figures \ref*{fig:human_yerkes_dodson}--\ref*{fig:friendship_compare}; 
Appendix \ref*{apx:tables}. Tables \ref*{tab:standized_results_a}--\ref*{tab:all_anova_restuls_b}; 

\section*{Acknowledgment}
This research was supported by the National Science Foundation under Grant No. IIS-1846221. Any opinions, findings, and conclusions or recommendations expressed in this material are those of the author(s) and do not necessarily reflect the views of the National Science Foundation.
\updated{The authors thank the associate editor for their insightful suggestions and anonymous reviewers for their critical comments, which have greatly helped to improve the quality of this paper.}

\ifCLASSOPTIONcaptionsoff
  \newpage
\fi


\vspace{-10pt}
\bibliography{all-biblatex}
\bibliographystyle{IEEEtran}

%

\begin{IEEEbiography}[{\includegraphics[width=1in,height=1.25in,clip,keepaspectratio]{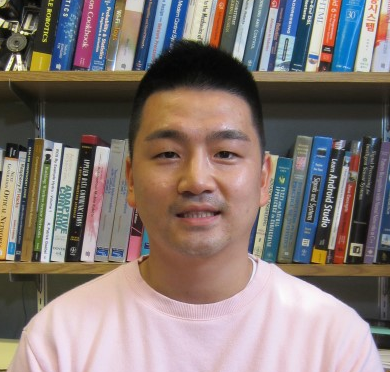}}]{Wonse Jo} received the B.S. in robotics engineering from Hoseo University, South Korea in 2013 and M.S. degrees in electronic engineering from the Kyung-Hee University, South Korea, in 2015. He is currently pursuing the Ph.D. degree in computer and information technology at Purdue University, West Lafayette, IN, USA. His research interests include social human-robot interaction (sHRI), affective robotics, human multi-robot interaction, robot perception, environmental robotics, and field robotics.\end{IEEEbiography}

\begin{IEEEbiography}[{\includegraphics[width=1in,height=1.25in,clip,keepaspectratio]{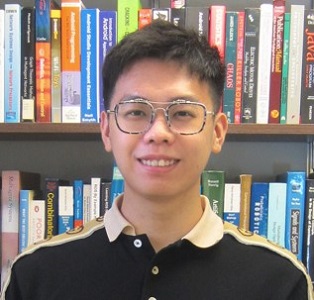}}]{Ruiqi Wang} (Student Member, IEEE) received a B.E. degree in robotics engineering from Beijing University of Chemical Technology, Beijing, China, in 2020. He is currently working towards a Ph.D. degree in the Department of Computer and Information Technology at Purdue University, West Lafayette, IN, USA. His research interests include human-robot interaction, affective robotics, and human-in-the-loop robot learning.\end{IEEEbiography}

\begin{IEEEbiography}[{\includegraphics[width=1in,height=1.25in,clip,keepaspectratio]{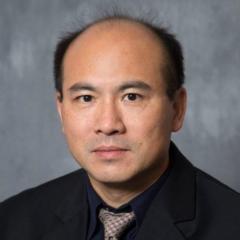}}]{Baijian Yang} (Member, IEEE) received his Ph.D. in Computer Science from Michigan State University, and his MS and BS in Automation (EECS) from Tsinghua University. He is currently a Professor at the Department of Computer and Information Technology, Purdue University. He served as a steering committee member of IEEE Cybersecurity Initiative between 2015 and 2017, and served as a board director of ATMAE from 2014-2016. His research interests include cybersecurity, data-driven security analytics, and applied machine learning. \end{IEEEbiography}


\begin{IEEEbiography}[{\includegraphics[width=1in,height=1.25in,clip,keepaspectratio]{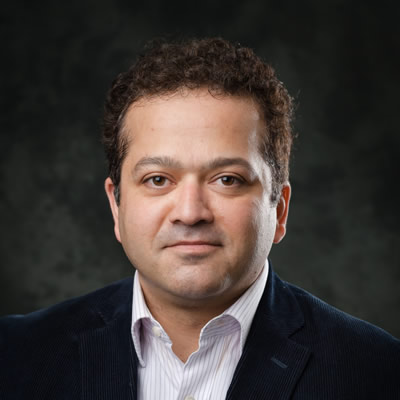}}]{Mo Rastgaar} (Senior Member, IEEE) received the B.S. degree from the Sharif University of Technology, Tehran, Iran, in 1995; the M.S. degree from Tehran Polytechnic, Iran, in 1998; and the Ph.D. degree from Virginia Polytechnic Institute and State University, Blacksburg, VA, USA, in 2008, all in mechanical engineering. 
He was a Post-Doctoral Associate in the Newman Laboratory for Biomechanics and Human Rehabilitation, Massachusetts Institute of Technology, Cambridge, MA, USA from 2008- to 2010. From 2011 to 2018, he was an assistant professor and associate professor at Michigan Tech, Houghton, MI, USA. In 2019, he joined Purdue University, West Lafayette, IN, USA, where he is currently an Associate Professor at Polytechnic Institute and the Director of the Human-Interactive Robotics Lab.\end{IEEEbiography}

\begin{IEEEbiography}[{\includegraphics[width=1in,height=1.25in,clip,keepaspectratio]{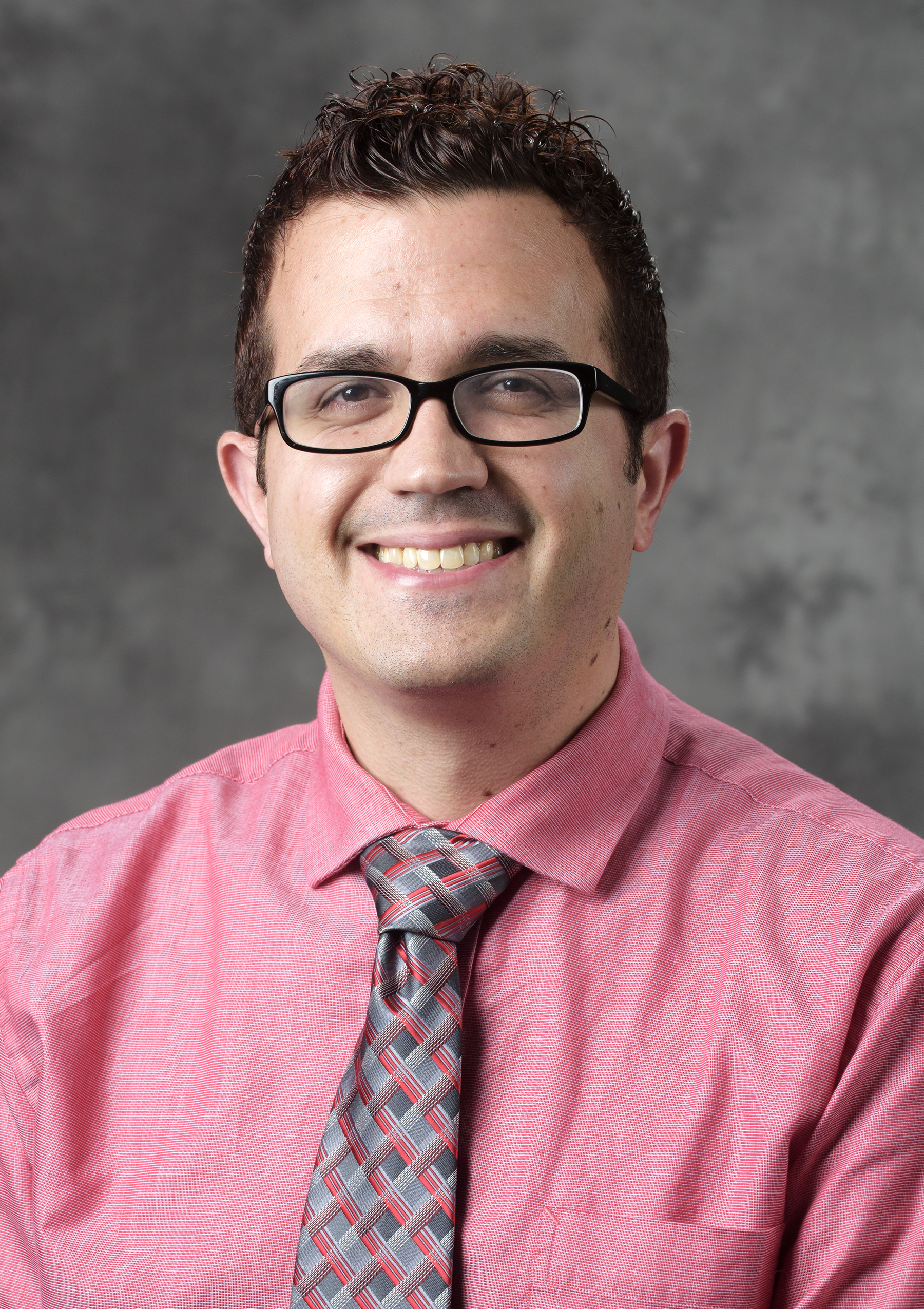}}]{Daniel Foti} received a B.A. degree in Biomedical Engineering from Harvard University in 2006. He completed his graduate studies at Stony Brook University, receiving his M.A. in Psychology in 2008 and his Ph.D. in Clinical Psychology in 2013. He completed his predoctoral clinical internship at McLean Hospital in Belmont, MA. He joined the faculty in the Department of Psychological Sciences at Purdue University as an Assistant Professor in 2013. He was promoted to Associate Professor with tenure in 2018. His research interests include using psychophysiological measures to study cognition, emotion, and reward, as well as abnormalities in these processes that are associated with vulnerability to psychiatric illnesses. \end{IEEEbiography}

\begin{IEEEbiography}[{\includegraphics[width=1in,height=1.25in,clip,keepaspectratio]{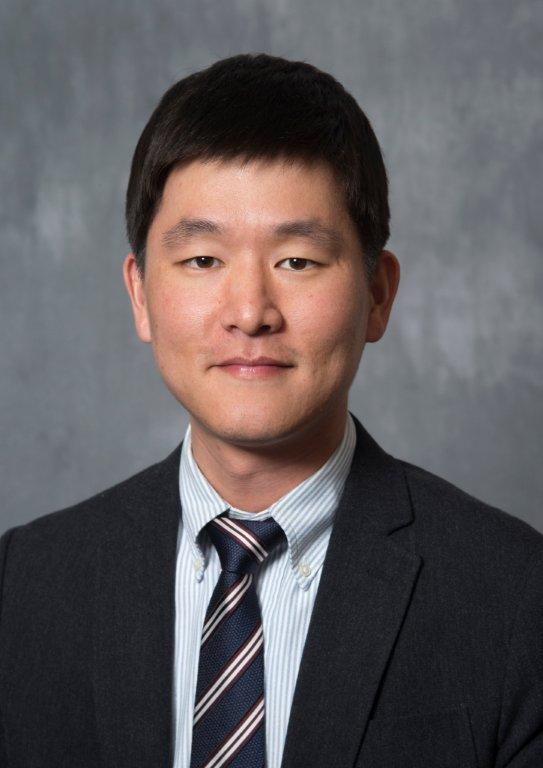}}]{Byung-Cheol Min} (Senior Member, IEEE) received a B.S. degree in electronics engineering and a M.S. degree in electronics and radio engineering from Kyung Hee University, Yongin, South Korea, in 2008 and 2010, respectively, and a Ph.D. degree in computer and information technology from Purdue University, West Lafayette, IN, USA, in 2014. He is currently an Associate Professor and University Faculty Scholar with the Department of Computer and Information Technology and the Director of the SMART Laboratory, Purdue University. Prior to this position, he was a Postdoctoral Fellow with the Robotics Institute, Carnegie Mellon University, Pittsburgh, PA, USA. His research interests include multi-robot systems, human–robot interaction, robot learning, with applications in field robotics, and assistive technology and robotics. \\He was a recipient of the NSF CAREER Award, in 2019; the Purdue PPI Outstanding Faculty in Discovery Award, in 2019; the Purdue CIT Outstanding Graduate Mentor Award, in 2019; the Purdue Focus Award, in 2020; the Purdue PPI Interdisciplinary Research Collaboration Award, in 2021; the Purdue Corps of Engagement Award, in 2022. He was named a Purdue University Faculty Scholar, in 2021. \end{IEEEbiography}




\newpage
\clearpage
\setcounter{page}{1}

\onecolumn
\textit{   }
\vspace{1.5cm}
\begin{center}
    \LARGE{Supplementary Materials for} \\
    \vspace{0.2cm}
    \Large{\textbf{Affective Workload Allocation for Multi-human Multi-robot Teams}} \\
    \vspace{0.3cm}
    \normalsize{Wonse Jo, Ruiqi Wang, Baijian Yang, Dan Foti, Mo Rastgaar, and Byung-Cheol Min}\\
    \vspace{0.5cm}
    Corresponding author: Byung-Cheol Min, \url{minb@purdue.edu} \\
    
    \vspace{0.3cm}
    \textit{IEEE Transactions on Human-Machine Systems} \\
    DOI: \url{10.1109/THMS.2023.XXXXX} \\
    \vspace{2cm}
    
\end{center}

\large{\textbf{The PDF file includes:}}
\begin{itemize}
\setlength{\itemindent}{.5in}
    \item Appendix \ref{apx:video}: Supplementary Video \& Website
    \item Appendix \ref{apx:part_req}: Participant Requirements
    \item Appendix \ref{apx:algorithm}: Algorithm \ref{alg:drl_env_gym}
    \item Appendix \ref{apx:post_interview}: Extended Post-interview Analysis
    \item Appendix \ref{apx:figures}: Figures \ref{fig:human_yerkes_dodson} to \ref{fig:friendship_compare}
    \item Appendix \ref{apx:tables}: Tables \ref{tab:standized_results_a} to \ref{tab:all_anova_restuls_b}

\end{itemize}

\twocolumn
\appendices


\section{Supplementary Video \& Website}\label{apx:video}
\setcounter{table}{0}
\begin{itemize}
    \item Team-based user experiment video: \url{https://youtu.be/qrmAQqfdLZk}
    \item Paper website: \url{https://sites.google.com/view/affective-workload-allocation}

\end{itemize}

\section{Participant Requirement}\label{apx:part_req}
\renewcommand{\thefigure}{S\arabic{figure}}
\setcounter{figure}{0}

For the team-based user experiment, we recruited 32 participants through flyers, social networking services, and email. All participants met the following requirements:

\begin{itemize} 
    \item Over 18 years old, 
    \item No medical history of brain disorders (e.g., stroke, brain tumor, surgery), no mental illness (e.g., depression, bipolar), 
    \item No heart diseases (e.g., high/low blood pressure, myocardial infarction), no vision or muscle impairment, 
    \item No skin irritation or allergic reaction to glycerin and saline fluids.
\end{itemize}

\section{Supplementary Algorithm}\label{apx:algorithm}
\renewcommand{\thealgorithm}{S\arabic{algorithm}}

\begin{algorithm}[h!]
    \caption{The DRL learning environment.}
    \label{alg:drl_env_gym}
    
    \begin{algorithmic}
        \State Initialize state observation $ \mathcal{S} \gets {(s^{o}_{1}, \dots, s^{o}_{n}), (s^{s}_{1}, \dots, s^{s}_{n})}$;
        \State Initialize actions space $\mathcal{A} \gets (a_{1}, \dots, a_{n})$
        \State Initialize weights $w \gets (\alpha_{p}, \beta_{p})$;
        \While{not terminated}
            \State Receive initial state observation $(S^{o}, S^{s})$ with random;
            \State Receive initial actions space $\mathcal{A}$;
            \ForAll{$episode$ = 1, $\dots$, i}
                \State Initial actions space $\mathcal{A}$ with random;
                \State Calculate reward $r(s, a)$;
                \State Calculate transition in actions space $\mathcal{A}^{\prime}$;
                \State Calculate $P_{team}^{\prime}$ according to $ \mathcal{S}$, $ \mathcal{A}^{\prime}$, and $w$;
                \State Calculate reward $r(s^{\prime}, a^{\prime})$;
                \If{Check termination criteria}:
                    \State Stop this episode;
                \Else
                    \State Update current state observation $s$;
                    \State $episode$ = $episode + 1$
                \EndIf
            \EndFor
            \State Updated policy $\pi$
        \EndWhile
    \end{algorithmic}
    
\end{algorithm}

\newpage

\section{Extended Post-interview Analysis}\label{apx:post_interview}
\renewcommand{\thetable}{S\arabic{figure}}
\setcounter{figure}{0}

We conducted post-interviews with the participants to gather feedback on our AWAC system and found that the majority of participants were satisfied with our AWAC system, which automatically changes the workload. Specifically, some participants reported that IS and AS were helpful in conducting missions:

\renewcommand{\labelitemi}{}
\begin{itemize}
    \item ``\textit{Personally, IS and AS were very helpful because I felt like I could lead the system} [P\_A of T4].''
\end{itemize}

On the other hand, other participants preferred changing the workload for the CCTV monitoring task rather than a fixed workload:
\begin{itemize}
    \item ``\textit{It was so cool that this system automatically recommended a new workload based on our performance! Honestly, I am not sure how accurate the system is, but I could definitely feel that the recommended workload was better than sticking to a fixed workload. I felt that it improved my working ability} [P\_B of T10].''
\end{itemize}

We also found unexpected results that some participants felt pressure from their teammates due to the mission scoreboard that was displayed on the CCTV monitoring program at the end of each task. We observed some participants quickly closed the scoreboard program to avoid seeing their results:

\begin{itemize}
    \item ``\textit{Wow! You (teammate) did a good job. How to get the scores? Do you have any strategies to get more scores? Next time, I will get more score than you (teammate)} [P\_B of T3].''
\end{itemize}

Interestingly, we observed that some participants felt sorry for getting lower mission scores compared to their teammates, especially if they were friends:

\begin{itemize}
    \item ``\textit{I am very sorry to my teammate. My personal score is lower than theirs (teammate)} [P\_A of T1].''
\end{itemize}

Furthermore, some participants requested to pause the missions as they experienced headaches or discomfort caused by the EEG headset. This is a known issue when using wearable biosensors, as evidenced by one participant who said;

\begin{itemize}
    \item ``\textit{Can you stop the experiment for a minute? I have a headache} [P\_A of T15].''
\end{itemize}

The participants were given a 5-minute break during which they removed their EEG headsets. After the break, their headaches disappeared, and they were able to continue with the remaining tasks in the experiment.

\renewcommand{\labelitemi}{-}

\section{Supplementary Figures}\label{apx:figures}
\renewcommand{\thetable}{S\arabic{figure}}
\setcounter{figure}{0}

\begin{figure}[ht!]
    \centering
    \includegraphics[width=0.9\linewidth]{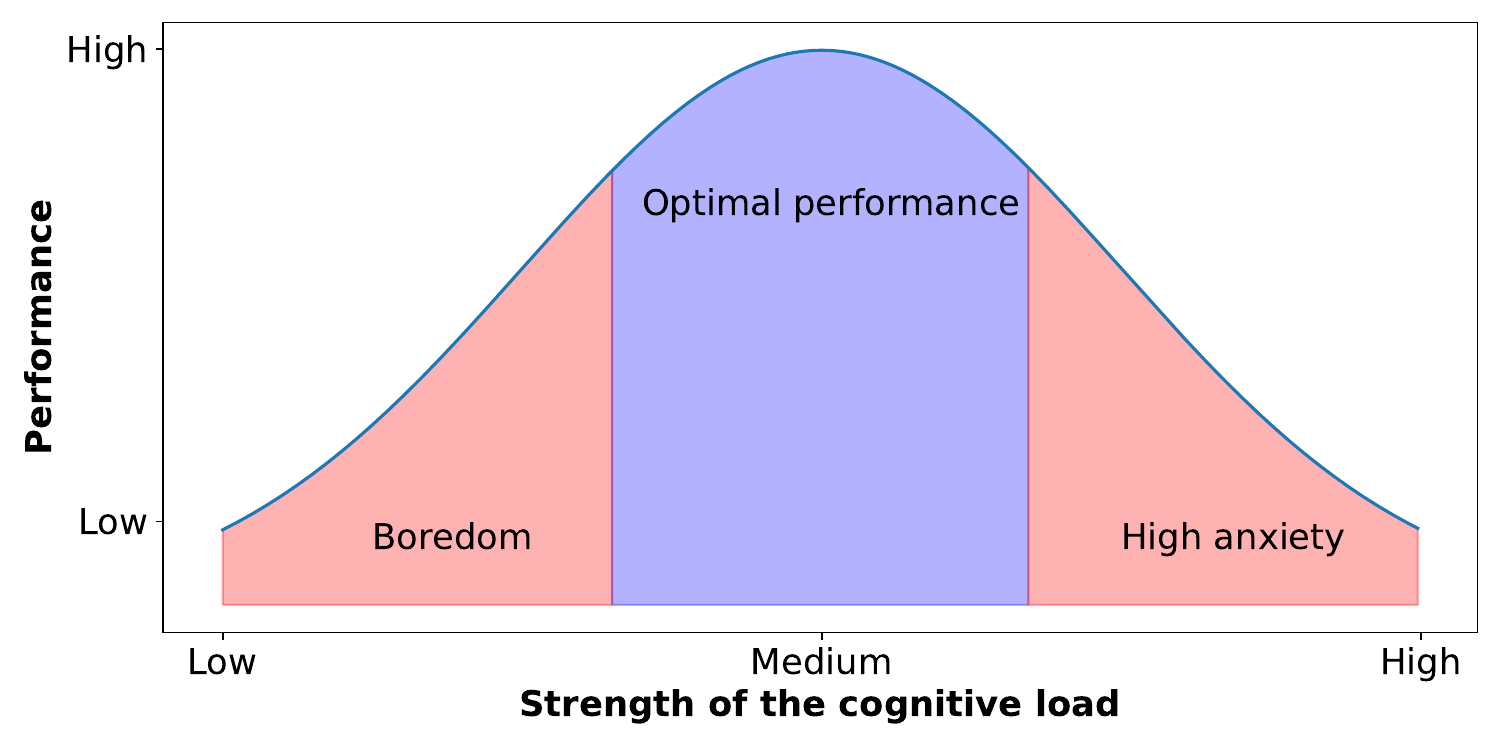}
    \caption{Generalized correlation between performance and levels of CWL based on the Yerkes-Dodson law \cite{yerkes1908relation}.}
    \label{fig:human_yerkes_dodson}
\end{figure}

\begin{figure}[ht!]
    \centering
    \begin{subfigure}[b]{0.8\linewidth}
        \centering 
        \includegraphics[width=1\linewidth]{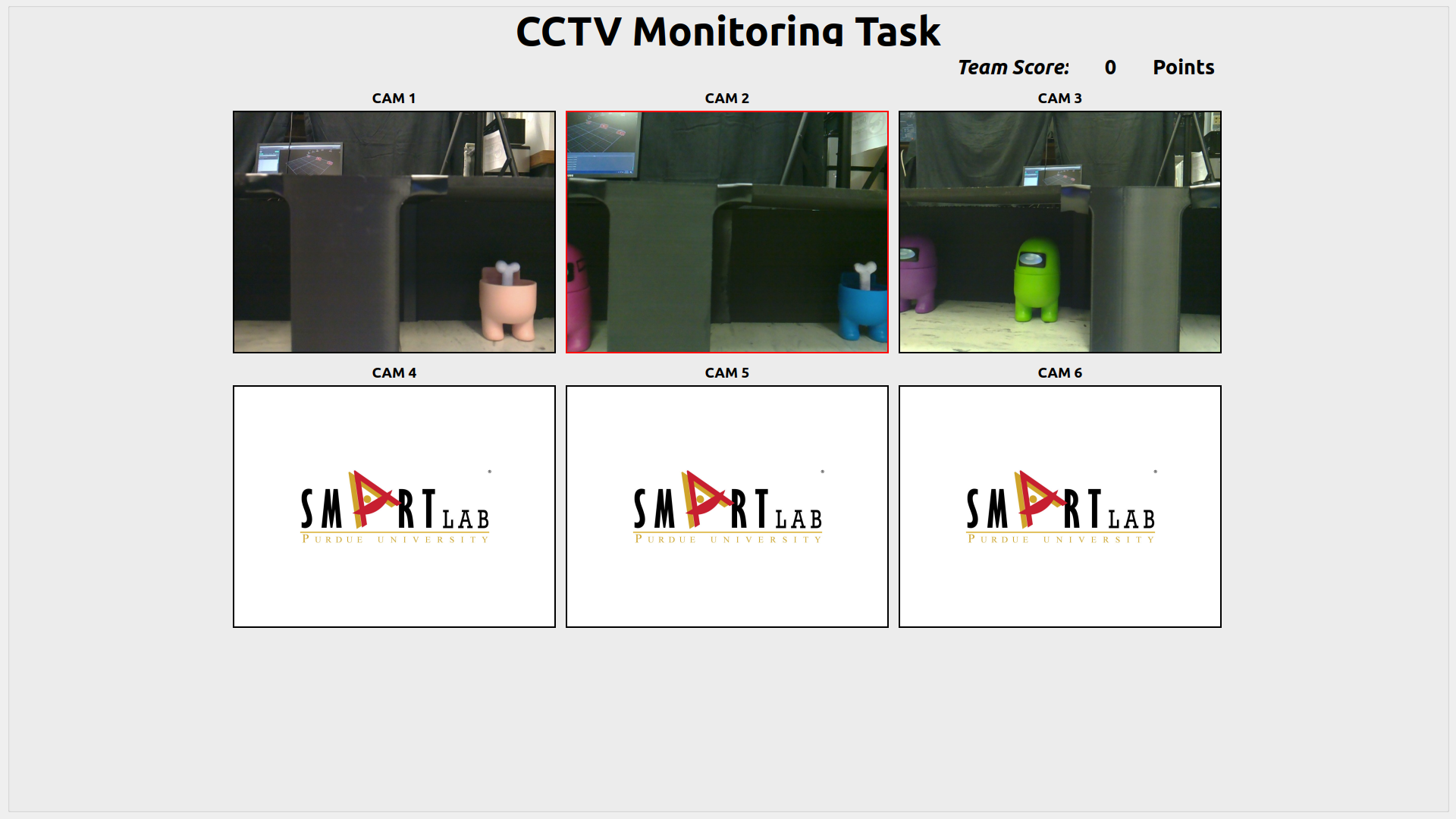}
        \caption{}
        \label{img:gui_start}
    \end{subfigure}    
    \hspace{3pt}
    
    \begin{subfigure}[b]{0.8\linewidth}
        \centering 
        \includegraphics[width=1\linewidth]{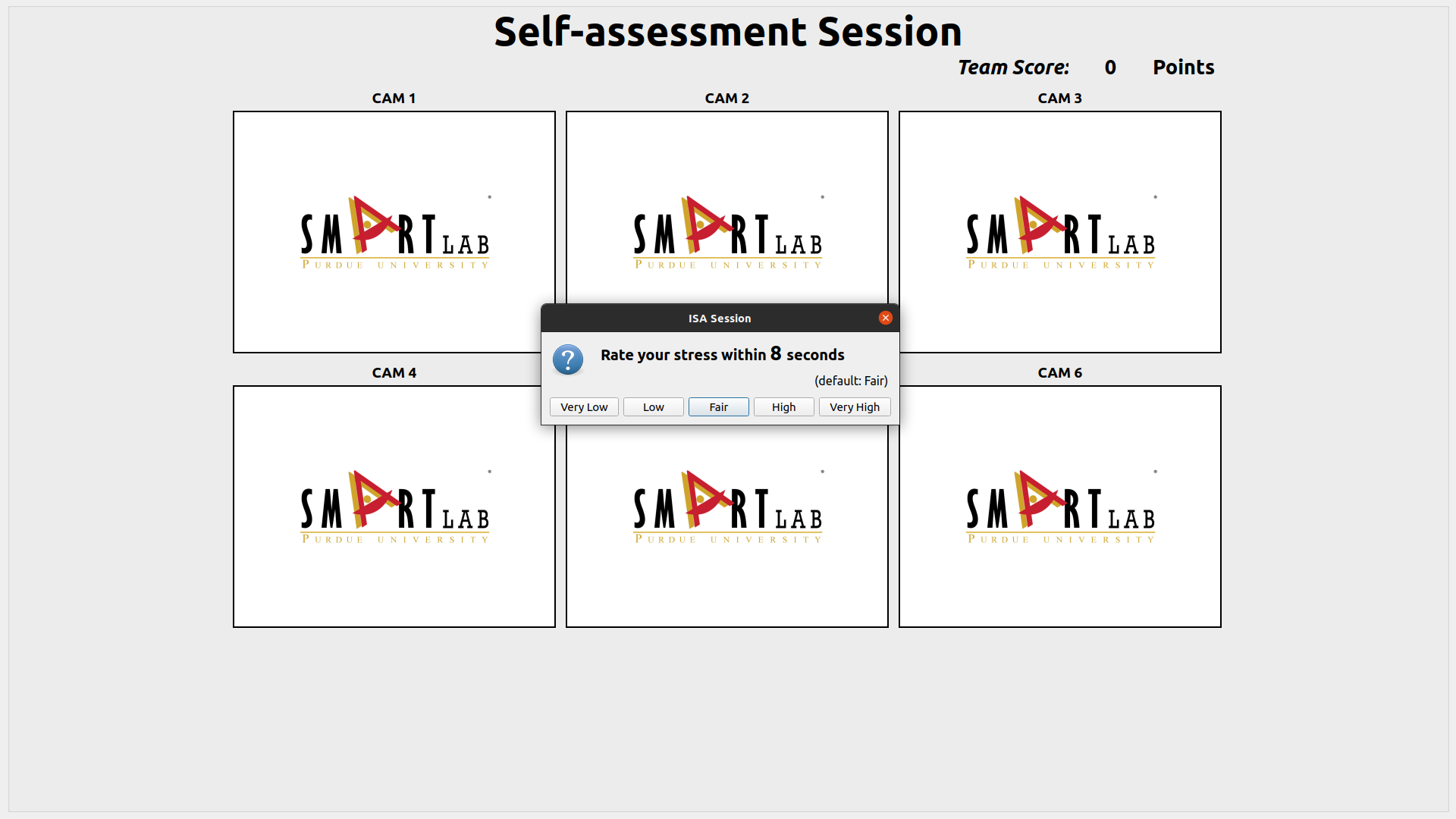}
        \caption{}
        \label{img:gui_isa}
    \end{subfigure}   
    
    \begin{subfigure}[b]{0.8\linewidth}
        \centering 
        \includegraphics[width=1\linewidth]{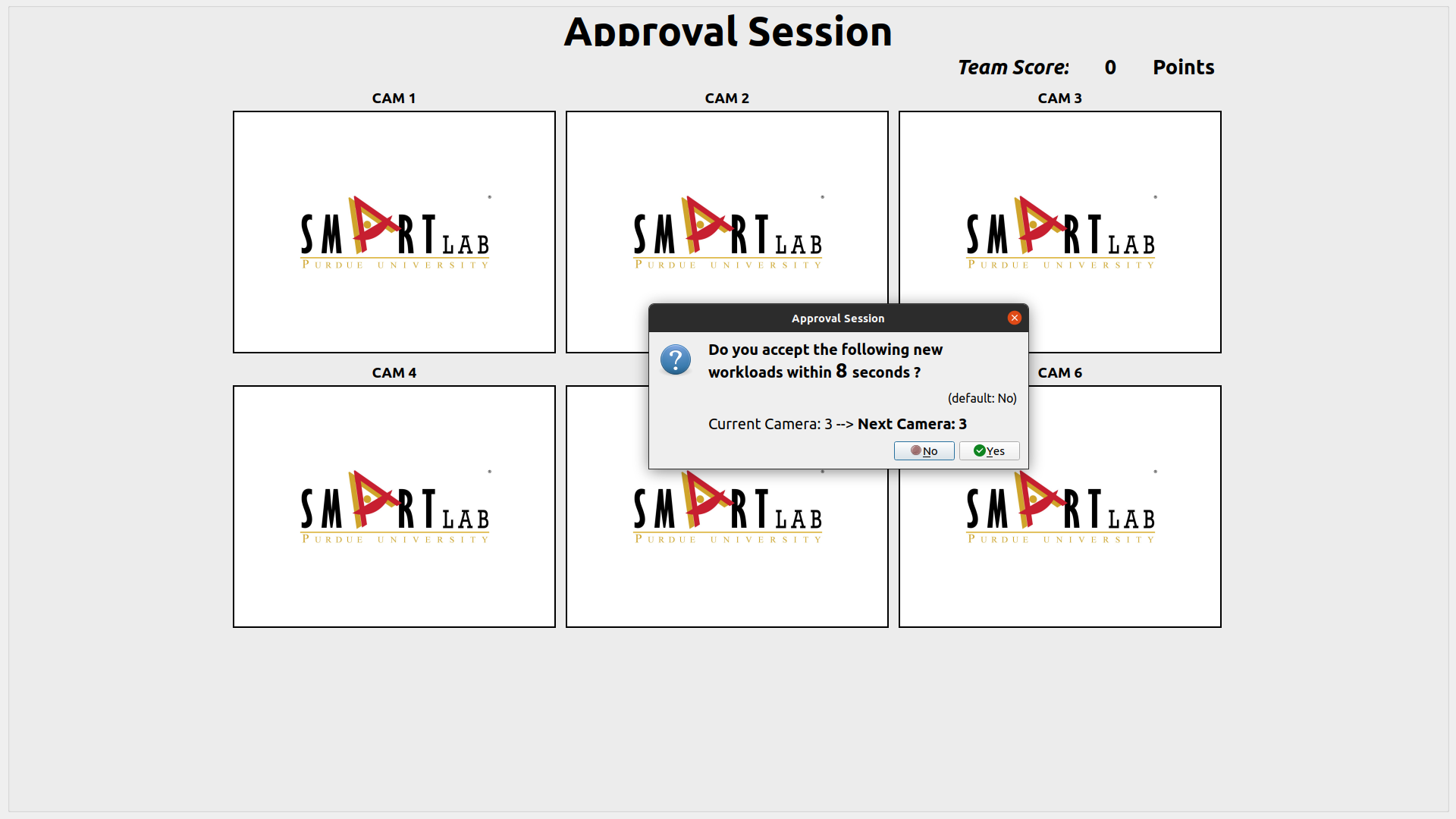}
        \caption{}
        \label{img:gui_approval}
    \end{subfigure}
    \caption{Graphic user interface (GUI) programs for human operators in the generalized surveillance scenario. (a) Main GUI for performing closed-circuit television (CCTV) monitoring tasks, (b) ISA session (IS) for participants to report current CWL by rating from very low (-2) to very high (2), and (c) Approval session (AS) for admitting transmitted additional workload from a teammate.}
    \label{fig:surveillance_GUIs} 
\end{figure}

\begin{figure}[ht!]
    \centering
    \includegraphics[width=0.83\linewidth]{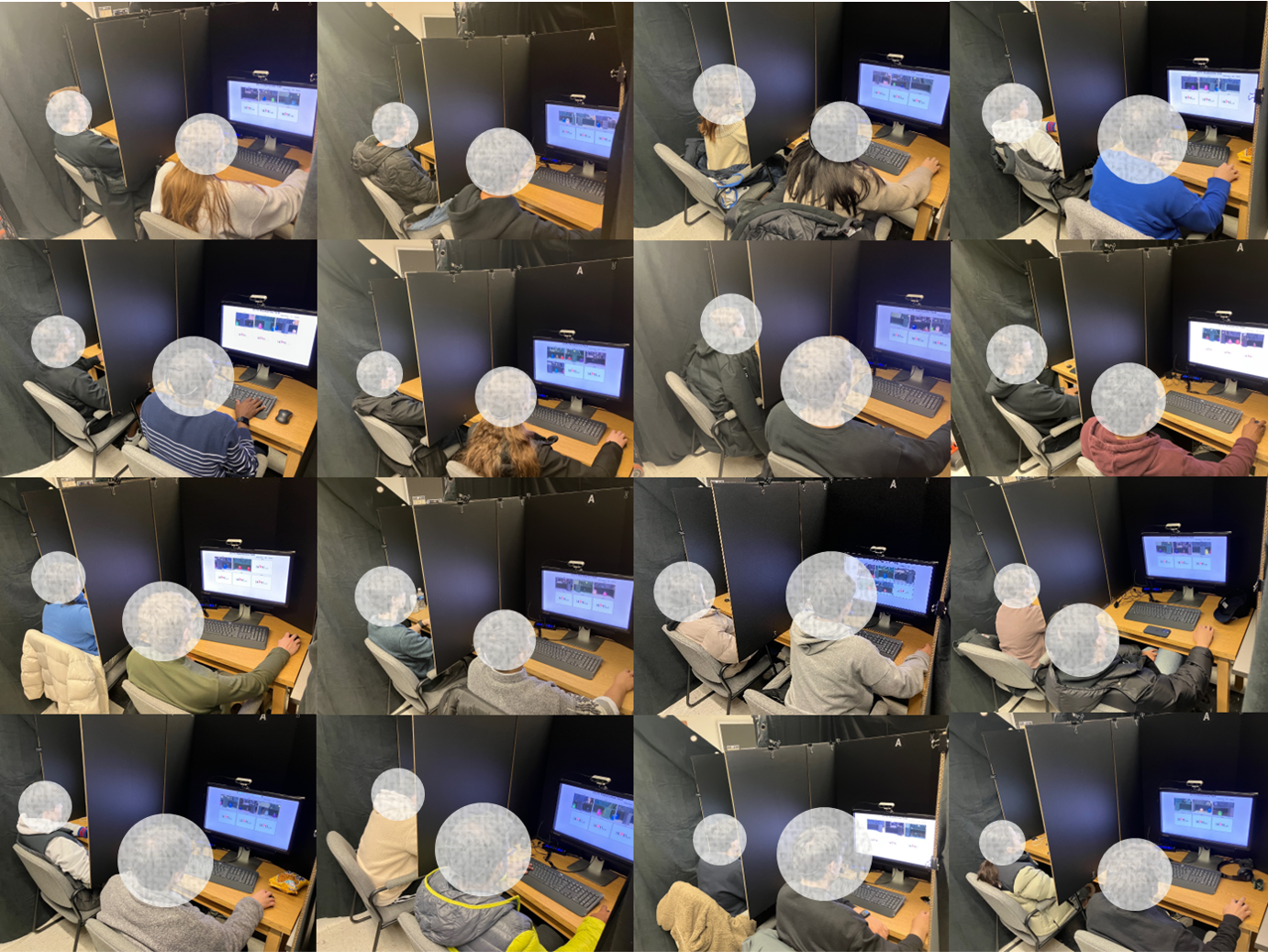}
    \caption{Experiment snapshots for our user experiments.}
    \label{fig:all_teams_study} 
\end{figure}

\begin{figure}[ht!]
    \centering 
    \includegraphics[width=1\linewidth]{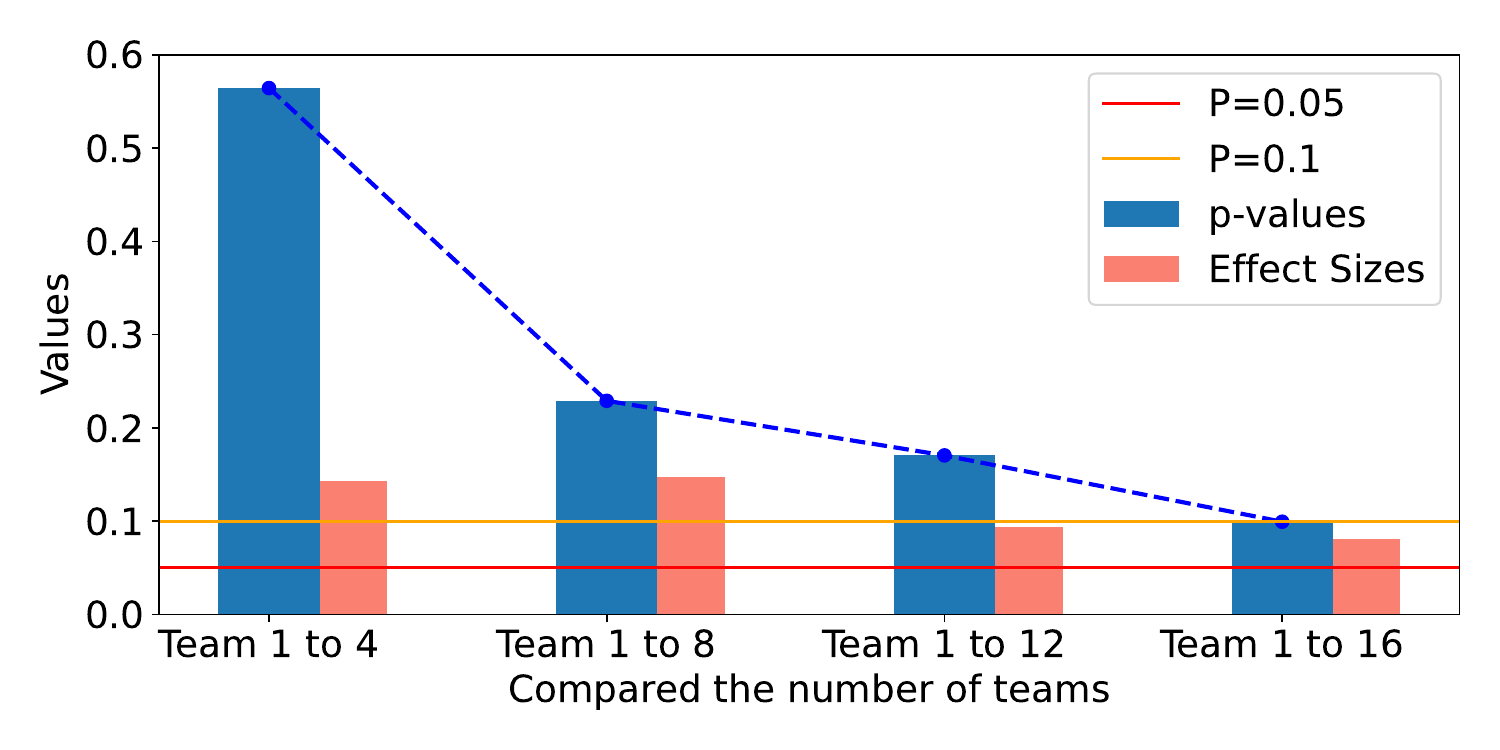}
    \caption{A trend of p-value according to the number of the team-based experiment.}
    \label{fig:p_treand} 
\end{figure}

\begin{figure}[ht!]
    \centering
    \includegraphics[width=1\linewidth]{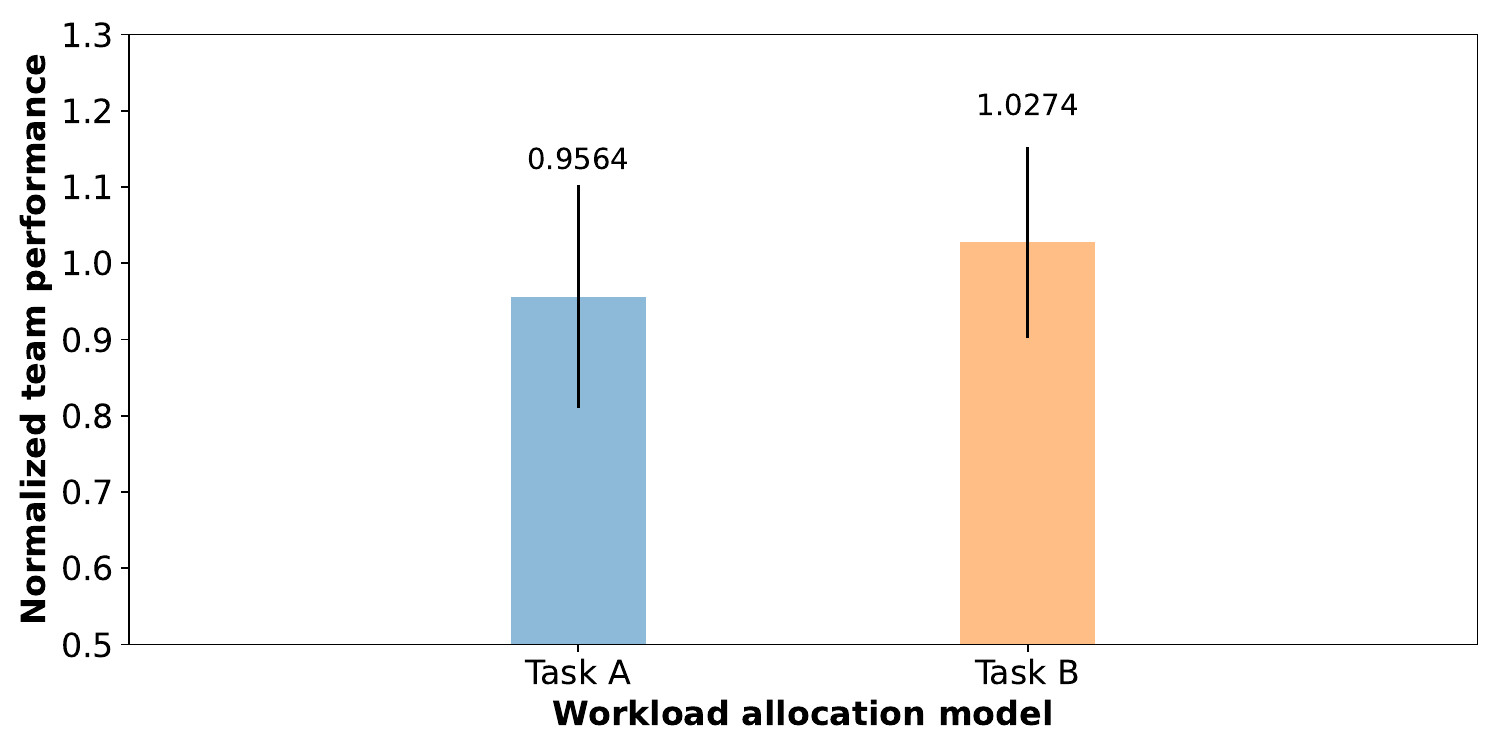}
    \caption{Distribution of team performance on Tasks A and B.}
    \label{fig:comparison_a_b}

\end{figure}

\begin{figure}[ht!]
    \centering
    \includegraphics[width=1\linewidth]{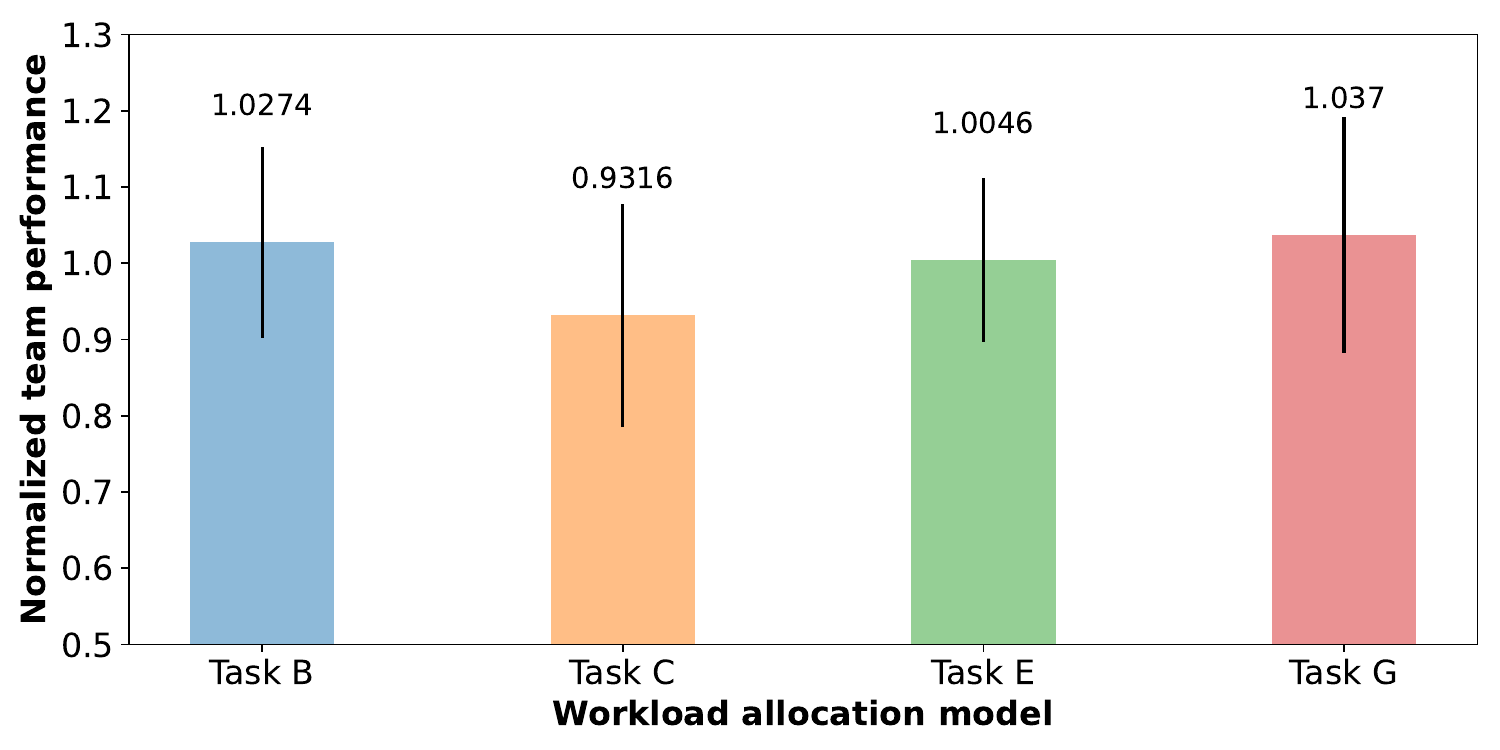}
    \caption{Distribution of team performance on four tasks (Tasks B, C, E, and G).}
    \label{fig:workload_method_exp}
\end{figure}

\begin{figure}[ht!]
    \centering
    \includegraphics[width=1\linewidth]{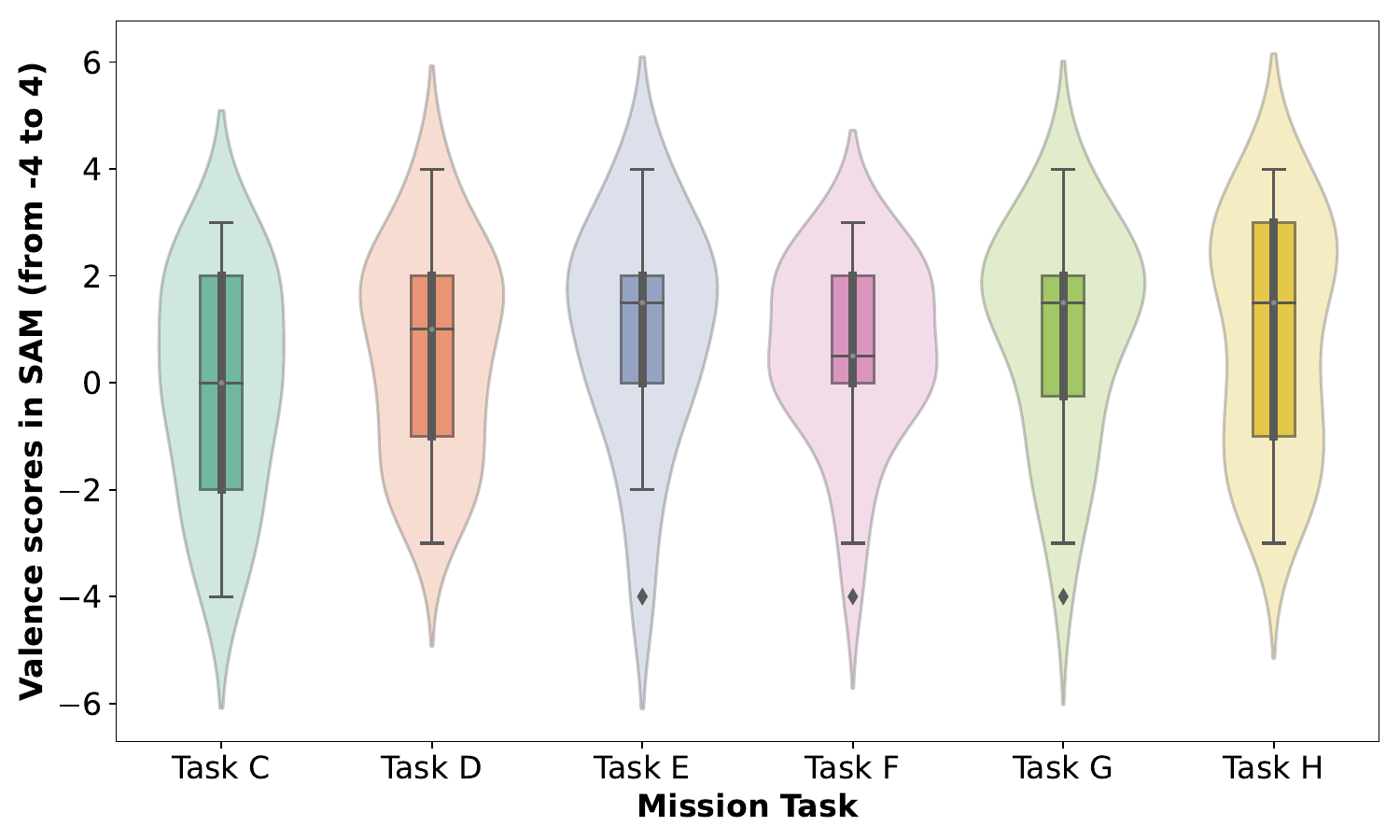}
    \caption{Valence ratings on the self-assessment manikin (SAM) for tasks applied to our AWAC.}
    \label{fig:sam_score} 
\end{figure}

\begin{figure}[ht!]
    \centering 
    \includegraphics[width=1\linewidth]{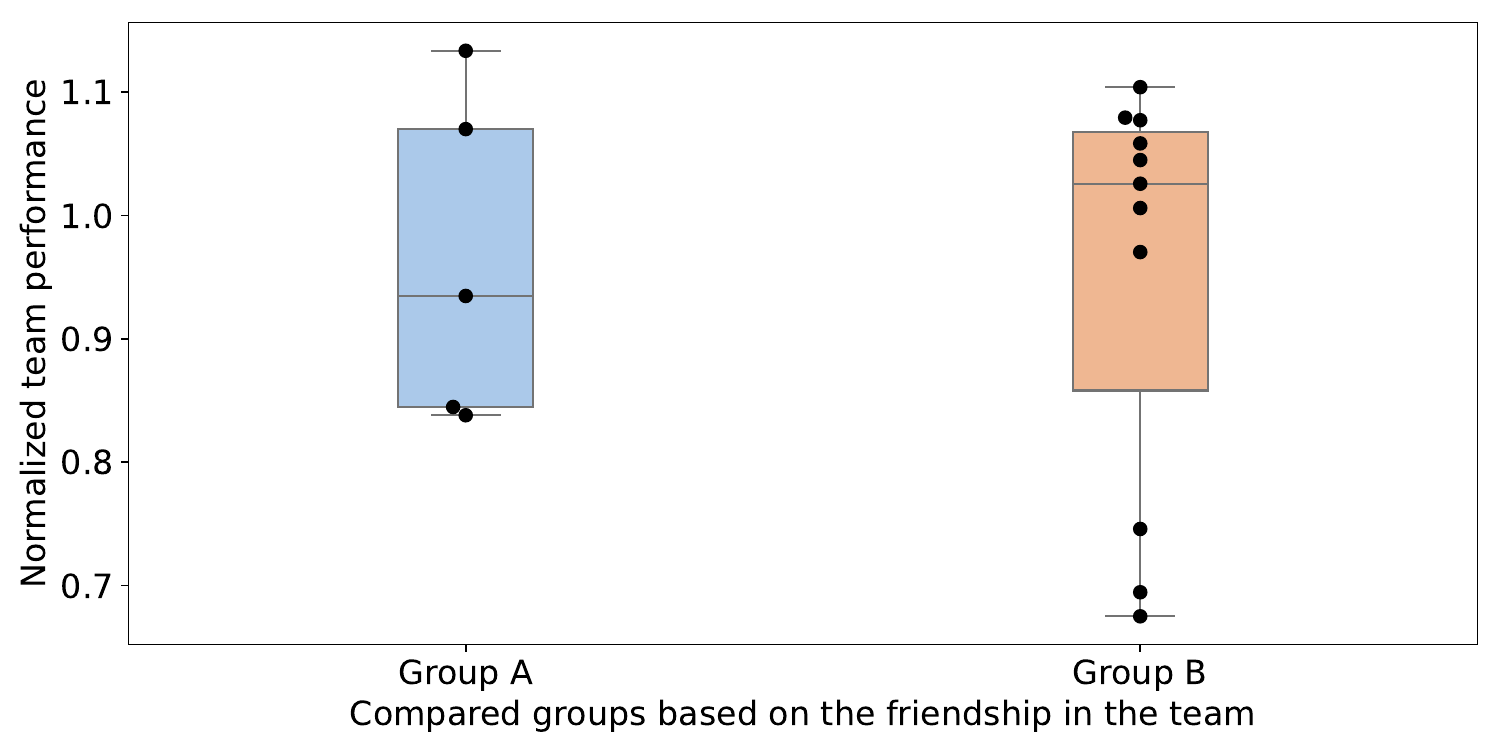}
    \caption{Comparison of team performance between Group A (Teams 1, 7, 8, 9, and 10) and Group B (Teams 2, 3, 4, 5, 6, 11, 12, 13, 14, 15, and 16) with and without friendship between two human subjects, respectively.}
    \label{fig:friendship_compare}
\end{figure}


\section{Supplementary Tables}\label{apx:tables}
\renewcommand{\thetable}{S\arabic{table}}
\setcounter{table}{0}
\begin{table}[h]
    \centering
    \caption{Normalized team performance scores of all 16 teams in Task A, D, F, and H.}
    \label{tab:standized_results_a}
    \resizebox{0.7\linewidth}{!}{%
        \begin{tabular}{|c||c|c|c|c|}
        \hline
        \rowcolor[HTML]{C0C0C0} 
        \textbf{} & \textbf{Task~A} & \textbf{Task~D} & \textbf{Task~F} & \textbf{Task~H}  \\ \hline \hline
        \textbf{T1} & 0.8447 & 1.0014 & 1.4295 & 0.9059  \\ \hline
        \textbf{T2} & 0.9703 & 0.7378 & 0.9110 & 0.9514  \\ \hline
        \textbf{T3} & 0.6945 & 1.0103 & 0.7546 & 0.9201  \\ \hline
        \textbf{T4} & 1.0060 & 1.0060 & 1.1039 & 0.9530  \\ \hline
        \textbf{T5} & 0.6750 & 1.0083 & 0.9356 & 1.0724 \\ \hline
        \textbf{T6} & 0.7458 & 0.7002 & 1.1509 & 0.9148 \\ \hline
        \textbf{T7} & 1.0700 & 0.8767 & 0.9778 & 0.9531 \\ \hline
        \textbf{T8} & 0.9347 & 1.0304 & 0.9375 & 0.8883 \\ \hline
        \textbf{T9} & 0.8380 & 1.1639 & 1.0658 & 1.1957 \\ \hline
        \textbf{T10} & 1.1335 & 1.0059 & 1.1573 & 1.2166 \\ \hline
        \textbf{T11} & 1.1040 & 1.1040 & 0.9189 & 1.0316 \\ \hline
        \textbf{T12} & 1.0772 & 0.9299 & 1.2296 & 1.0875 \\ \hline
        \textbf{T13} & 1.0257 & 0.9441 & 0.9996 & 1.1074 \\ \hline
        \textbf{T14} & 1.0449 & 0.8652 & 0.9012 & 0.9758 \\ \hline
        \textbf{T15} & 1.0793 & 0.9580 & 1.1122 & 1.1273 \\ \hline
        \textbf{T16} & 1.0584 & 1.0894 & 1.1868 & 1.1846 \\ \hline \hline
        \textbf{Mean} & 0.9564	& 0.9645	& 1.0483	& 1.0303 \\ \hline
        \textbf{S.D.} & 0.1507	& 0.1234	& 0.1639	& 0.1121 \\ \hline
        \end{tabular}%
}
\end{table}

\begin{table}[h]
    \centering
    \caption{Results of the rmANOVA test for the team performance scores obtained in Task A, D, F, and H.}
    \label{tab:all_anova_restuls_a}
    \resizebox{1\columnwidth}{!}{%
    \begin{tabular}{|l|l|l|l|l|l|}
        \hline
        \rowcolor[HTML]{C0C0C0} 
        \textbf{Source} & \textbf{DF} & \textbf{Sum of Square} & \textbf{Mean Square} & \textbf{$F$ Statistic} & \textbf{$p$-value} \\ \hline \hline
        \textbf{Between groups} & 3 & 0.103	  & 0.034	 &  2.214 & 0.0995 \\ \hline
        \textbf{Within groups} & 60 & 1.161  & 0.019	 &  &  \\ \hline
        \textbf{Error} & 45 & 0.6955 & 0.016	 &  &  \\ \hline
        \end{tabular}%
    }
\end{table}

\begin{table}[h]
    \centering
     \caption{Normalized team performance scores of all 16 teams for the team-based user experiment in Task B, C, E, and G.}
    \label{tab:standized_results_b}
    \resizebox{0.7\linewidth}{!}{%
        \begin{tabular}{|c||c|c|c|c|}
        \hline
        \rowcolor[HTML]{C0C0C0} 
        \textbf{} & \textbf{Task~B} & \textbf{Task~C} & \textbf{Task~E} & \textbf{Task~G} \\ \hline \hline
        \textbf{T1} & 1.0855 & 0.8524 & 0.9556 & 0.9250 \\ \hline
        \textbf{T2} & 1.0605 & 0.9561 & 1.2503 & 1.1625 \\ \hline
        \textbf{T3} & 1.1658 & 1.0981 & 1.1633 & 1.1934 \\ \hline
        \textbf{T4} & 0.9265 & 0.9080 & 1.0245 & 1.0721 \\ \hline
        \textbf{T5} & 1.1877 & 0.7883 & 1.0531 & 1.2796 \\ \hline
        \textbf{T6} & 1.0355 & 1.1885 & 1.1348 & 1.1294 \\ \hline
        \textbf{T7} & 1.0992 & 0.9531 & 0.9621 & 1.1082 \\ \hline
        \textbf{T8} & 1.0878 & 1.0796 & 0.9101 & 1.1315 \\ \hline
        \textbf{T9} & 0.8012 & 0.7032 & 1.0977 & 1.1345 \\ \hline
        \textbf{T10} & 0.9911 & 0.5994 & 0.9614 & 0.9347 \\ \hline
        \textbf{T11} & 0.9976 & 0.9380 & 0.8870 & 1.0189 \\ \hline
        \textbf{T12} & 0.7724 & 0.9868 & 0.9816 & 0.9351 \\ \hline
        \textbf{T13} & 0.8559 & 0.9767 & 0.9996 & 1.0911 \\ \hline
        \textbf{T14} & 1.1776 & 1.0919 & 0.9261 & 1.0173 \\ \hline
        \textbf{T15} & 1.1299 & 0.9302 & 0.8291 & 0.8341 \\ \hline
        \textbf{T16} & 1.0650 & 0.8547 & 0.9366 & 0.6244 \\ \hline \hline
        \textbf{Mean} & 1.0275 & 	0.9316 & 1.0046	& 1.0370 \\ \hline
        \textbf{S.D.} & 0.1293	& 0.1513	& 0.1110	& 0.1593 \\ \hline
        \end{tabular}%
}
\end{table}


\begin{table}[!h]
    \centering
    \caption{Results of the rmANOVA test for the team performance scores obtained in Task B, C, E, and G.}
    \label{tab:all_anova_restuls_b}
    \resizebox{1\columnwidth}{!}{%
    \begin{tabular}{|l|l|l|l|l|l|}
        \hline
        \rowcolor[HTML]{C0C0C0} 
        \textbf{Source} & \textbf{DF} & \textbf{Sum of Square} & \textbf{Mean Square} & \textbf{$F$ Statistic} & \textbf{$p$-value} \\ \hline \hline
        \textbf{Between groups} & 3 & 0.1092	  & 0.0364	 &  2.3588 & 0.0842 \\ \hline
        \textbf{Within groups} & 60 & 1.1596  & 0.0193	 &  &  \\ \hline
        \textbf{Error}          & 45 & 0.6946 & 0.0154	 &  &  \\ \hline
        \end{tabular}%
    }
\end{table}
\newpage
\vspace{100pt}


\end{document}